\documentclass[10pt,twocolumn,letterpaper]{article}

\usepackage{cvpr}              

\usepackage{graphicx}
\usepackage{amsmath,amssymb}
\usepackage{multirow}
\usepackage{url}
\usepackage{algorithm}
\usepackage{amsmath}
\usepackage{amsfonts}
\usepackage{algorithmic}
\usepackage{caption}
\usepackage{bm}

\usepackage[dvipsnames]{xcolor}

\newcommand{\para}[1]{\vspace{.05in}\noindent\textbf{#1}\quad}

\definecolor{cvprblue}{rgb}{0.21,0.49,0.74}
\usepackage[pagebackref,breaklinks,colorlinks,allcolors=cvprblue]{hyperref}

\graphicspath{{./Imgs/}}
\DeclareGraphicsExtensions{.pdf,.jpg,.png}

\def\paperID{7043} 
\def\confName{CVPR}
\def\confYear{2025}

\title{Exploiting Temporal State Space Sharing for Video Semantic Segmentation}

\author{%
Syed Ariff Syed Hesham\textsuperscript{1,2}, 
Yun Liu\textsuperscript{3*}, 
Guolei Sun\textsuperscript{4}\thanks{Corresponding authors: Yun Liu and Guolei Sun}, 
Henghui Ding\textsuperscript{5}, 
Jing Yang\textsuperscript{6}, \\ 
Ender Konukoglu\textsuperscript{4},
Xue Geng\textsuperscript{2}, 
Xudong Jiang\textsuperscript{1}  \\
\textsuperscript{1} Nanyang Technological University \quad
\textsuperscript{2} Institute for Infocomm Research, A*STAR \\
\textsuperscript{3} VCIP, CS, Nankai University \quad
\textsuperscript{4} CVL, ETH Zurich \quad
\textsuperscript{5} Fudan University \quad
\textsuperscript{6} Guizhou University
}

\begin{document}
\maketitle

\begin{abstract}
Video semantic segmentation (VSS) plays a vital role in understanding the temporal evolution of scenes. Traditional methods often segment videos frame-by-frame or in a short temporal window, leading to limited temporal context, redundant computations, and heavy memory requirements. To this end, we introduce a \textbf{Temporal Video State Space Sharing (TV3S)} architecture to leverage Mamba state space models for temporal feature sharing. Our model features a selective gating mechanism that efficiently propagates relevant information across video frames, eliminating the need for a memory-heavy feature pool. By processing spatial patches independently and incorporating shifted operation, TV3S supports highly parallel computation in both training and inference stages, which reduces the delay in sequential state space processing and improves the scalability for long video sequences. Moreover, TV3S incorporates information from prior frames during inference, achieving long-range temporal coherence and superior adaptability to extended sequences. Evaluations on the VSPW and Cityscapes datasets reveal that our approach outperforms current state-of-the-art methods, establishing a new standard for VSS with consistent results across long video sequences. By achieving a good balance between accuracy and efficiency, TV3S shows a significant advancement in spatiotemporal modeling, paving the way for efficient video analysis. The code is publicly available at \url{https://github.com/Ashesham/TV3S.git}.
\end{abstract}

\setlength\belowcaptionskip{+2ex}
\begin{figure}[!t]
    \centering
    \includegraphics[width=\linewidth]{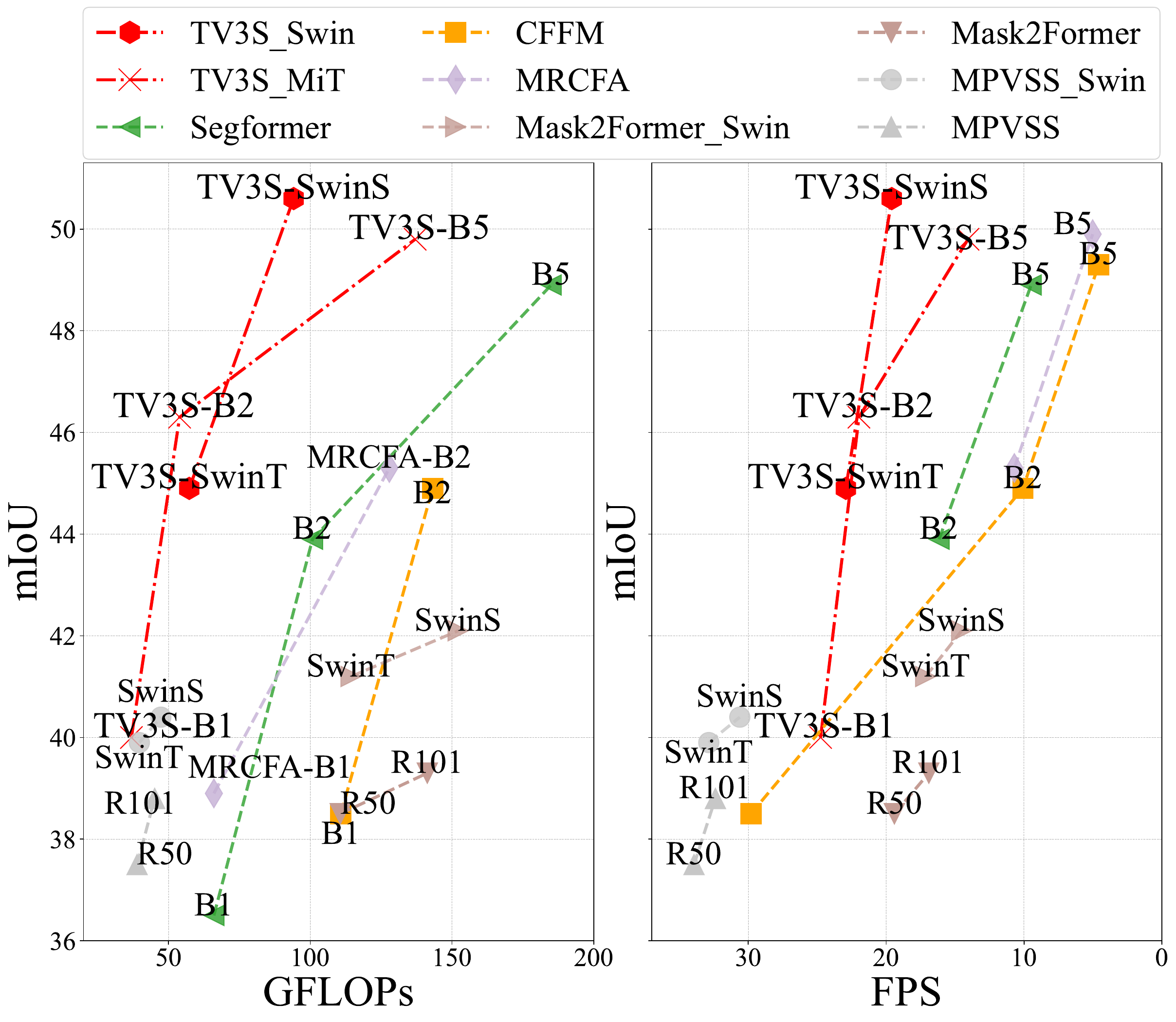}
    \caption{Comparison of the proposed TV3S with baseline models for VSS. By enhancing temporal information, our TV3S demonstrates superior performance over the baselines.}
    \label{fig:res_comp}
    \vspace{-.1in}
\end{figure}

\section{Introduction}
\label{sec:intro}
Semantic Segmentation has achieved substantial progress through the introduction of convolutional neural networks \cite{FCN, mobilenetv2, DeepLab, ronneberger2015u, PSPNet} and, more recently with vision transformers \cite{ViT,Swin,DETR} due to their ability to capture spatial patterns and contextual information. However, these techniques are mainly designed for static images and do not leverage the temporal dynamics present in videos
\cite{CFFM,MPVSS,MOSE,MeViS,guo2018visual,atluri2018spatio}.

Recently, research attention has increasingly shifted toward video semantic segmentation (VSS) \cite{siam2021video,siam2018comparative,litjens2017survey,tsakanikas2018video}, due to its potential to benefit a wide range of practical applications.
Unlike static image segmentation, VSS requires models to track motion, adapt to changes in appearance, and handle interactions among objects across consecutive frames. These challenges necessitate sophisticated techniques to accurately interpret dynamic environments where objects may occlude one another, shift positions, or exhibit varying states of motion and stillness \cite{qian2024controllable,Mask2Former}.

The contemporary approaches of VSS fall into two main categories. The first leverages \textit{optical flow}, modeling pixel movement between consecutive frames to align features and support object tracking, thereby enhancing temporal coherence. These methods effectively propagate temporal information, making them useful for fine-grained temporal alignment \cite{MPVSS,gao2023exploit,sevilla2016optical,ding2020every,cheng2017segflow}. However, they have significant drawbacks, including high computational costs due to the complexity of estimating accurate flow fields, especially in scenes with occlusions or sudden changes. Additionally, optical flow relies on precise motion estimation, where inaccuracies in dynamic scenes can propagate and lead to reduced accuracy in video segmentation.

The second category involves \textit{feature aggregation}, where information from multiple frames are combined to improve segmentation accuracy \cite{xu2012streaming, kundu2016feature}. The researchers relied on recurrent neural networks (RNNs) particularly Convolutional Long Short-Term Memory (ConvLSTM) \cite{shi2015convolutional} to introduce temporal structures to the models \cite{valipour2017recurrent,emre2017semantic,nabavi2018future,pfeuffer2019semantic,wang2021noisy}. Recently, VM-RNN \cite{tang2024vmrnn} has advanced this area by applying an efficient LSTM model to work with Vision Mamba \cite{zhu2024vision} to capture spatiotemporal dynamics. While these methods can capture both spatial and temporal information, they face challenges when scaling to handle long video sequences, largely due to the high computational and memory demands of recurrent networks. Furthermore, the sensitivity of LSTMs to sequence length often leads to training instability, requiring careful tuning and optimization \cite{wang2021noisy,shi2015convolutional,pfeuffer2019semantic}. These limitations underscore the need for more scalable architectures.

Due to such challenges in past feature aggregation methods, the focus in this domain has shifted towards the usage of transformer-based models \cite{CFFM,MRCFA,CFFM++,hu2020temporally,wang2021temporal,li2019attention,li2021video,lao2023simultaneously} \cite{vaswani2017attention}. However, these approaches still face challenges, including high memory costs and limited scalability, particularly with long video sequences or high-resolution frames. 
Efficient information sharing across numerous frames continues to be a challenge, making real-time applications difficult.

To address the challenges of integrating spatial and temporal information in VSS, we present \textbf{Temporal State Space Sharing (TV3S)}, designed to overcome the limitations of existing models. Unlike traditional optical flow methods, which struggle with inaccuracies in dynamic scenes, and RNN- and Transformer-based architectures, which are constrained by expensive feature pooling and high computational costs, TV3S takes a more efficient approach by processing spatial patches in parallel while dynamically sharing temporal information across video frames. Specifically, a spatially encoded input frame is split into discrete patches, which are processed independently through a series of TV3S blocks, each containing \textbf{Temporal State Space (TSS)} modules with a selective gating mechanism that effectively integrates and propagates spatiotemporal information across frames, ensuring minimal computational overhead. To further enhance motion handling at the edges of patches, we introduce a shifted window-based approach that works with un-shifted and shifted encoded features, enabling the model to capture movements near the boundaries while still maintaining efficient temporal information sharing. These components work jointly to integrate both local and long-range temporal features, significantly improving VSS performance while avoiding excessive computational costs and improving overall efficiency as evident in \cref{fig:res_comp}.

In summary, the contributions of our paper include: 
\begin{itemize}

    \item We present the Temporal Video State Space Sharing (TV3S) architecture, a novel framework that shares and propagates temporal information across video frames efficiently and effectively. 

    \item We process spatial patches independently with a selective gating mechanism efficiently, thus enabling parallel computation during both training and inference and supporting scalability for long, high-resolution video sequences.

    \item We design a shifted window-based approach within the TV3S block, enhancing temporal state space sharing and capturing long-range spatial context effectively.

\end{itemize}
Through extensive experiments on the VSPW \cite{VSPW_dataset} and Cityscapes \cite{cordts2016cityscapes} datasets, we demonstrate that our approach surpasses existing state-of-the-art methods, establishing new benchmarks for efficiency and accuracy in VSS.

\setlength\belowcaptionskip{-1ex}
\begin{figure*}[!t]
    \centering
    \includegraphics[width=\linewidth]{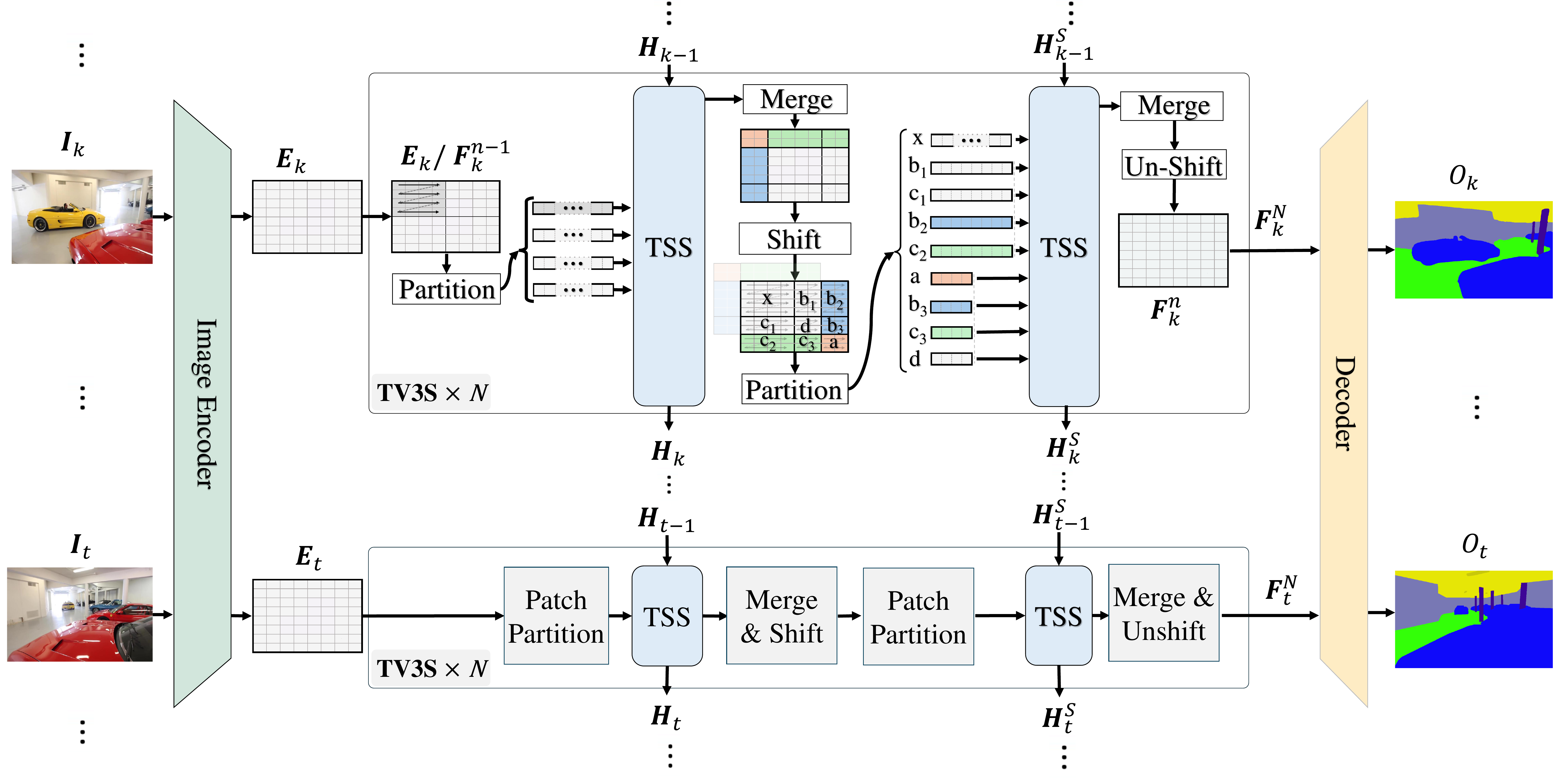}
    \caption{Overview of the proposed TV3S architecture, illustrating the encoder-decoder framework that employs state space models and our TSS module based on Mamba \cite{mamba_paper} for independent spatial and temporal processing.}
    \label{fig:ourmodel}
    \vspace{-.1in}
\end{figure*}

\section{Related Work}
\label{sec:related}

Semantic segmentation began with natural images, with the introduction of fully convolutional networks (FCNs) \cite{FCN} pioneering an end-to-end pixel-wise classification framework. Building on this foundation, subsequent works enhanced segmentation accuracy by adopting atrous convolutional layers \cite{DeepLab,DeepLabv3+}, employing pyramid architectures \cite{PSPNet,DeepLab,DeepLabv3+}, leveraging encoder-decoder architectures \cite{ronneberger2015u,CCL,o2015learning}, and incorporating attention mechanisms \cite{wang2018non,fu2019dual,hu2018squeeze}. More recently, transformer-based architectures like SegFormer \cite{xie2021Segformer}, Segmenter \cite{Segmenter}, SETR \cite{SETR}, and Mask2Former \cite{Mask2Former}, have further advanced the field by learning global dependencies. While these developments have significantly improved image semantic segmentation, transitioning them to VSS introduces new challenges related to temporal coherence and efficient exploitation of temporal information.

\subsection{VSS Based on Recurrent Neural Networks} 

Early works in this field explored recurrent neural networks (RNNs) for temporal modeling. Valipour \etal \cite{valipour2017recurrent} incorporated a recurrent unit between the encoder and decoder, significantly improving video segmentation performance. Evaluations by \cite{emre2017semantic} on different RNN structures such as ConvRNN \cite{srivastava2015unsupervised}, ConvGRU \cite{ballas2016delving}, and ConvLSTM \cite{shi2015convolutional} on the KITTI dataset \cite{geiger2013vision} demonstrated the superiority of ConvLSTM in handling video sequences. Further explorations have combined ConvGRU with optical flow to represent pixel displacements and maintain temporal continuity \cite{nilsson2018semantic}, while bidirectional ConvLSTM has been applied to merge temporally adjacent features, enhancing stability across frames \cite{nabavi2018future}. Despite these advancements, the high computational and memory cost of RNNs for processing longer video sequences poses notable challenges, highlighting the need for more efficient models capable of handling long-range temporal dependencies \cite{shi2015convolutional,pfeuffer2019semantic}.

\subsection{VSS Based on CNNs and Transformers}
To address RNN limitations, CNN-based methods were developed for temporal modeling. DFF \cite{zhu2017deep} and Accel \cite{Accel} utilized optical flow to propagate features between frames, reducing redundancy. ClockNet \cite{shelhamer2016clockwork} and LLVS \cite{li2018low} introduced adaptive feature reuse to exploit semantic similarity across frames for efficiency and temporal coherence.

Recognizing the limitations of optical flow and CNNs for long-range temporal modeling (see \cref{sec:intro}), recent advancements \cite{li2021video,li2022video,li2023tube,an2023temporal} have seen a shift towards using transformers \cite{vaswani2017attention} in VSS. Transformers, with their self-attention mechanism, can capture global dependencies across frames, making them suited for temporal feature aggregation. 
Among notable approaches, MPVSS \cite{MPVSS} proposes a memory-augmented transformer framework to capture multi-frame dependencies, enabling efficient temporal aggregation across longer video sequences. 
Similarly, CFFM \cite{CFFM} and MRCFA \cite{MRCFA} employ multi-resolution cross-frame attention to handle temporal variations by disentangling static and dynamic contexts within video frames, allowing for a refined segmentation process that distinguishes between stationary and moving elements.

Despite advancements, transformer-based methods face high computational costs due to the quadratic complexity of self-attention, especially with high-resolution frames or long videos \cite{vaswani2017attention}. Additionally, they are usually designed to learn temporal information in a short video sequence due to the high complexity.
This gap highlights the need for more efficient and holistic models that can simultaneously manage computational costs while effectively modeling long-range temporal information in VSS.

\subsection{State Space Models}
State space models (SSMs) present a promising alternative for temporal modeling, addressing the shortcomings of RNNs and transformers. Unlike RNNs, which suffer from scaling challenges with sequence length, SSMs like S4 \cite{gu2021efficiently,gu2021combining} exhibit with linear complexity by imposing diagonal structures on state matrices, making them more efficient for long data sequences. Enhanced through HiPPO \cite{gu2020hippo} initialization, SSMs handle extensive dependencies while requiring less memory, making them well-suited for tasks requiring to store long temporal context. Recently, Mamba \cite{mamba_paper} introduced a selective-scan mechanism to process temporal data efficiently. This advancement has spurred adaptations in vision-specific models, such as Vision Mamba \cite{zhu2024vision} and VMamba \cite{liu2024vmambavisualstatespace}, which incorporate Mamba blocks in a hierarchical structure to overcome directional sensitivities and maintain scalability across high-resolution inputs. Additionally, Vim \cite{zhu2024vision} refined the scanning techniques to prevent overfitting. Furthermore, U-Mamba \cite{ma2024u} has explored a hybrid network architecture that combines SSMs with convolutional layers. VideoMamba \cite{li2025videomamba} represents one of the early frameworks to leverage SSM-based modules for video understanding, focusing on video clip classification. However, it is limited by its offline processing, which increases memory and computational demands, restricts dense semantic segmentation, and hinders its ability to capture the global context of sequences.

However, despite these advancements, current vision-specific models largely rely on hierarchical designs fail to fully exploit varying temporal resolutions and context lengths, highlighting a critical need for improved mechanisms for temporal state sharing. To this end, our proposed TV3S architecture integrates parameter-efficient SSMs to enhance effective temporal state sharing, thus proving advantageous for VSS. This integration not only underscores the computational efficiency and scalability of SSMs in VSS but also emphasizes their suitability for tasks requiring long-range temporal coherence, representing a significant novelty in the field.

\setlength\belowcaptionskip{+2ex}
\begin{figure}[!t]
    \centering
    \includegraphics[width=\linewidth]{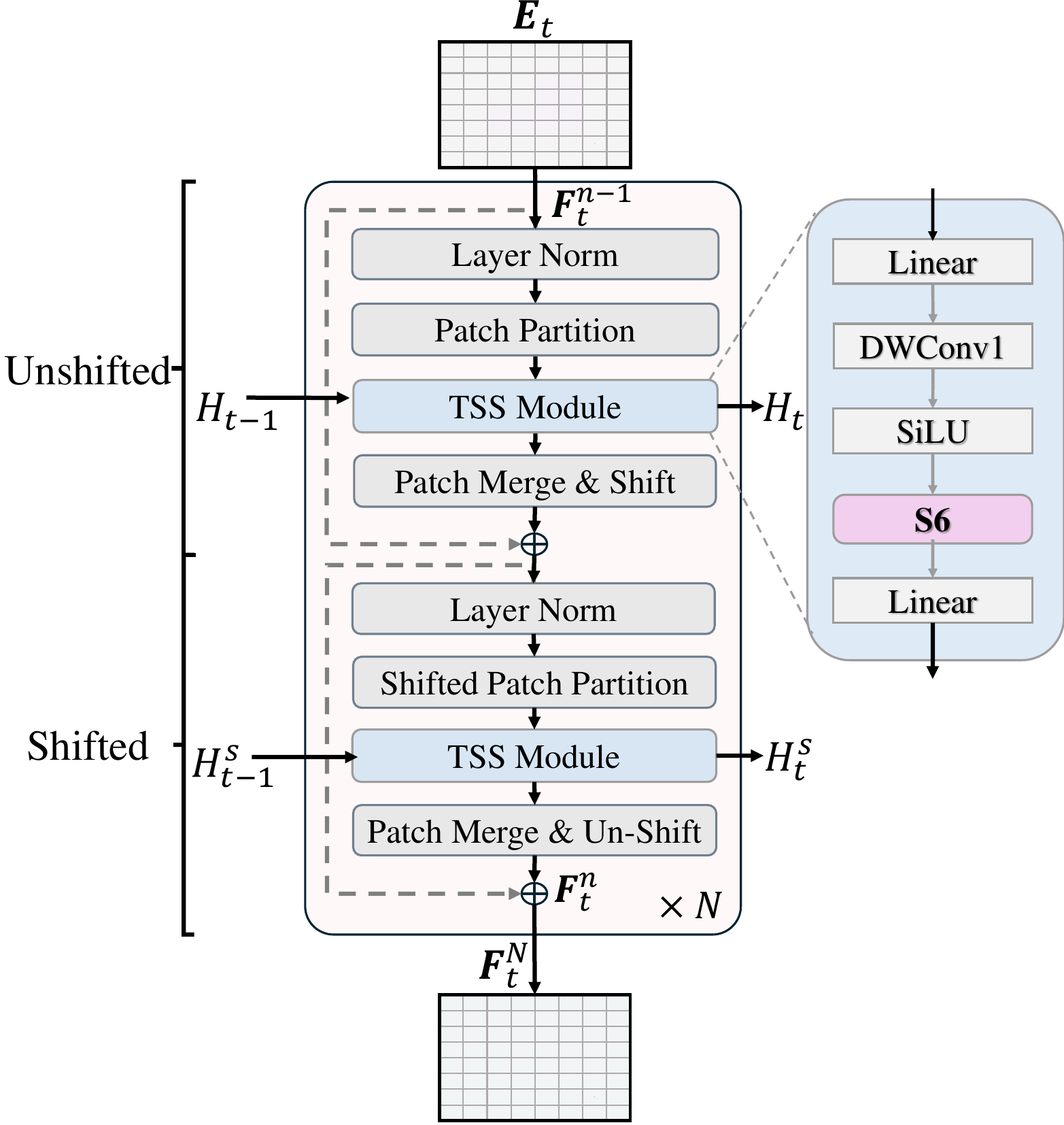}
    \vspace{0.1pt}
    \caption{Internal structure of the TV3S block, illustrating the flow of internal operations, and the propagation of hidden states for efficient spatiotemporal integration.}
    \label{fig:Block}
    \vspace{-.1in}
\end{figure}

\section{Methodology}
\label{sec:method}
\cref{fig:ourmodel} illustrates the proposed TV3S architecture, which effectively captures and integrates temporal dependencies within video sequences using a state space modeling framework. It aims to leverage both temporal and spatial information to achieve accurate and consistent segmentation results.

\subsection{Overall Architecture}
\label{ssec:overall_arch}

The architecture processes a sequence of video frames denoted as $\{\bm{I}_{t-l}, ..., \bm{I}_{t-k}, ..., \bm{I}_{t-1}, \bm{I}_t\}$, where $I_t$ represents the frame at time $t$, and $\{l, ..., k, ..., 1\}$ are temporal offsets capturing frames from the past relative to $t$. These frames are fed into an image encoder such as MiT~\cite{xie2021Segformer} and Swin~\cite{Swin}. The encoder produces a corresponding set of encoded feature $\{\bm{E}_{t-l}, ..., \bm{E}_{t-k},  ..., \bm{E}_{t-1},  \bm{E}_t\}$, capturing rich spatial context. 

The TV3S architecture processes these feature maps sequentially through a series of $N$ TV3S blocks (with a default value of $N=4$), each containing two TSS modules that independently process spatial patches of the feature map, facilitating parallelization. Layer normalization is applied before each TSS module, and residual connections are included after each module to stabilize training. Each TSS module consists of operations including a linear projection, a 1-dimension depthwise convolution (DWConv1), a SiLU activation feeding into the state space model (S6), and finally ending with another linear projection to project back to the input dimension, as detailed in \cref{fig:Block}.

\subsection{Temporal Video State Space Sharing Block}
\label{ssec:TV3S}

The TV3S block is the core unit of the TV3S architecture, which enables seamless integration of temporal information across frames. Each block consists of two TSS modules that employ a state space model to capture and propagate temporal dependencies within frames. The encoded feature map $\bm{E}_t$ is partitioned into non-overlapping patches of size $w \times w$. Each patch $\bm{P}^{i,j}_t$ is then flattened, transforming the spatial dimensions into a linear sequence suitable for parallel processing, with $i$ and $j$ representing the patch indices. Mathematically, $\bm{P}^{i,j}_t$ can be expressed as
\begin{gather}
\begin{aligned}
    \bm{P}^{i,j}_t &= \bm{E}_t[i:i+w,j:j+w,:] \in \mathbb{R}^{C_E\times w\times w}, \\
    \bm{\hat{P}}^{i,j}_t &= {\rm Flatten}(\bm{P}^{i,j}_t) \in \mathbb{R}^{C_E\times (w\times w)},
    \label{eq:Flatten}
\end{aligned}
\end{gather}
where $C_E$ denotes the number of channels in $\bm{E}_t$ after encoding $\bm{I}_t$. In contrast to traditional Mamba variants ~\cite{liu2024vmambavisualstatespace, zhu2024vision, ma2024u}, which process sequences sequentially, our approach parallelizes the handling of each window or patch, offering two main advantages. First, the encoder effectively encodes complex spatial information across channels, eliminating for further processing of spatial relationships. Second, minimal changes between consecutive frames allow window-based processing to efficiently capture temporal dynamics, enabling a highly parallelized and computationally efficient model.

\para{State space model for temporal aggregation.} 

Each TSS module within a TV3S block utilizes a state space model to capture temporal dependencies between frames. Given encoded feature map $\bm{E}_t$, the hidden state $\bm{H}_t$ is updated using the current flattened patch $\bm{\hat{P}}^{i,j}_t$ and the previous hidden state $\bm{H}_{t-1}$: 
\begin{gather}
\begin{aligned}
        \bm{H}_{t}^{i,j} &= f_A(\Delta,\bm{A}_s)\bm{H}_{t-1}^{i,j}+f_B(\Delta,\bm{A}_s,\bm{B}_s)\bm{\hat{P}}_{t}^{i,j}, \\ 
        \bm{F}_t^{i,j} \hspace{2.5mm} &= \bm{C}_s\bm{H}_{t}^{i,j}.
    \label{eq:1}
\end{aligned}
\end{gather}
Here, $\bm{A}_s \in \mathbb{R}^{N_s \times N_s}$, $\bm{B}_s \in \mathbb{R}^{N_s \times C_E}$, and $\bm{C}_s \in \mathbb{R}^{C_E \times N_s}$ are learnable state space matrices. As seen in \cref{eq:1}, the matrices $\bm{A}$ and $\bm{B}$ are discretized using the time scale parameter $\bm{\Delta}_s$ with fixed discretization functions $f_A(.)$ \& $f_B(.)$, referred as \textit{discretization rules}, as described in \cite{mamba_paper} due to their many advantages. Here, $N_s$ denotes the hidden states dimensionality of the state space model and the aggregated feature output is denoted as $\bm{F}_t^{i,j}$. This mechanism enables each patch to independently capture and propagate its temporal dependencies across two dimensions $(i, j)$.Consequently, temporal data storage consists of a total of $\frac{W}{w} \times \frac{H}{w}$ hidden states, promoting effective temporal consistency and high parallelization.

\para{Shifting and edge handling.} 
To enhance spatial context sharing, the reshaped feature maps undergo a shifting operation inspired by Swin Transformer \cite{Swin}. This shift is parameterized as $s \times s$, where $s < w$ (typically $s = w/2$). The shifting rearranges the blocks so that edge blocks receive information from adjacent regions, thereby enriching spatial-temporal representation. The shifted feature maps are then re-partitioned into patches, with those on the right and bottom edges further subdivided into smaller sub-blocks of dimensions $2 \times w \times \frac{w}{2}$, $2 \times \frac{w}{2} \times w$, and $4 \times \frac{w}{2} \times \frac{w}{2}$, as visualized in \cref{fig:ourmodel}. These re-partitioned patches are processed through a second TSS module in the same TV3S block, integrating the newly acquired spatial context with temporal information.

In the proposed architecture, each TV3S block consists of a pair of TSS modules, with one module processing unshifted blocks and the other handling shifted blocks. By stacking $N$ such TV3S blocks, the architecture forms a deep hierarchical structure termed TV3S architecture, which is capable of capturing intricate temporal dependencies across frames. The final aggregated feature representation, $\bm{F}^N$, is passed through a linear projection layer to map it to $C$ classes, followed by an interpolation to produce the final segmentation output.

\subsection{Training Strategy}
\label{ssec:Training}

The training strategy for the TV3S architecture focuses on optimizing both temporal feature aggregation and robust spatial feature extraction to enable efficient learning of short- and long-term dependencies within video sequences, facilitating high VSS performance.

During training, sequential input frames are taken at the intervals $\{\bm{I}_{t-9}, \bm{I}_{t-6}, \bm{I}_{t-3}, \bm{I}_t\}$ the same as \cite{VSPW_dataset}. Each of the frames in set \( \{\bm{I}_{t-k}\} \) with $k=\{9,6,3,0\}$, is passed through the encoder to generate encoded features \( \bm{E}_{t-k} \) encapsulating rich spatial information. An intermediate feature \( \hat{\bm{O}}_{t-k} \) is extracted straight from \( \bm{E}_{t-k} \) with the use of linear projection that aligns the channel dimensions of \( \bm{E}_{t-k} \) with the number of classes \( C \) in the dataset. Concurrently, the encoded features \( \bm{E}_{t-k}\) are fed into the TV3S blocks to integrate temporal information, producing aggregated features \( \bm{F}^N_{t-k} \).

With both output predictions extracted, a weighted cross-entropy between these output predictions $\bm{O}$ and the ground-truth segmentation masks $\bm{M}$ is computed as training loss for all frames in the input sequence. The training loss is computed as:
\begin{equation}
    \mathcal{L} = \lambda \sum_{k = \{9, 6, 3, 0\}} \mathcal{L}_{\text{CE}}(\hat{\bm{O}}_{t-k}, \bm{M}_{t-k}) + \mathcal{L}_{\text{CE}}(\bm{O}_t, \bm{M}_t).
    \label{eq:train_CE}
\end{equation}
Loss formulation presented in \cref{eq:train_CE} employs a dual-loss strategy to ensure accurate segmentation of the final frame \( \bm{O}_t \) while preserving spatial relationships in the intermediate features \( \hat{\bm{O}}_{t-k} \) from the encoder. By applying cross-entropy losses $\mathcal{L}_{\text{CE}}$, with intermediate losses weighted at $\lambda = 0.5$, the model effectively balances spatial feature learning and temporal coherence, enhancing its capability to extract temporally consistent spatial information for improved segmentation across consecutive frames.

\begin{algorithm}[!t]
    \caption{Inference Procedure for TV3S Architecture}
    \label{alg:inference}
    \begin{algorithmic}[1]
        \STATE \textbf{Input:} Frame sequence \( \{ \bm{I}_t \} \)
        \STATE \textbf{Output:} Segmentation map \( \bm{O}_t \)

        \STATE \textbf{Initialize:} Hidden states \( \bm{H}\) for all TV3S blocks
        \FOR{each frame \( \bm{I}_t \) in the sequence}
            \STATE \( \bm{E}_t \gets \text{Encoder}(\bm{I}_t) \quad \text{// Extract features} \)
            \FOR{each patch \( (i,j) \) in \( \bm{E}_t \)}
                \STATE \( \bm{\hat{P}}^{i,j} \gets \text{Flatten}(\bm{E}_t[i:i+w,j:j+w,:]) \)
                \FOR{each TV3S block \( n  \) from 1 to \( N\)}
                    \STATE \text{//Update Features and Hidden State.}
                    \STATE \(\bm{F}^{n,i,j}_t , \bm{H}_t^{n,i,j} \gets \text{TV3S}^n(\bm{\hat{P}}^{i,j}, \bm{H}^{n,i,j})\)
                \ENDFOR
            \ENDFOR
            \STATE \( \bm{O}_t \gets \text{Interpolate}(\text{Linear}(\bm{F}_t^N)) \quad \text{  // Segmentation} \)
            \STATE \( \bm{H} \gets \{\bm{H}_t\} \quad  \text{// Store for next frame} \)
        \ENDFOR
    \end{algorithmic}
\end{algorithm}

\subsection{Inference Procedure}
\label{ssec:Inference}

During inference, the TV3S architecture processes each frame sequentially while leveraging stored hidden states to maintain temporal coherence across the video sequence, as detailed in Algorithm \ref{alg:inference}. For each frame $\bm{I}_t$, the encoder first extracts the encoded features $\bm{E}_t$, which are then partitioned into non-overlapping spatial patches indexed by $(i,j)$ and flattened into $\bm{P}_{i,j}$. Each flattened patch is processed through the TV3S blocks, updating the current hidden states $\bm{H}_t$ using the previous states $\bm{H}_{t-1}$. After processing all patches, the aggregated feature map $\bm{F}_t^N$ is reconstructed from the updated hidden states. This feature map is then passed through a linear projection layer and interpolated to match the original image resolution, producing the final segmentation map $\bm{O}_t$. The updated hidden states $\bm{H}_t$ are then stored back in $\bm{H}$ for use with subsequent frames, ensuring continuous temporal integration and consistent segmentation results.

\setlength\belowcaptionskip{-1ex}
\begin{table*}[!t]
  \centering
  \setlength{\tabcolsep}{4.8mm}
  \resizebox{\linewidth}{!}{
  \begin{tabular}{l|c|c|c|c|c|c|c}\toprule
    Methods & Backbones & mIoU$\uparrow$ & mVC\textsubscript{8}$\uparrow$ & mVC\textsubscript{16}$\uparrow$ & GFLOPs$\downarrow$ & Params(M)$\downarrow$ & FPS$\uparrow$ \\
    \midrule
    Segformer\textsuperscript{\textit{$\dagger$}}  \cite{xie2021Segformer}& MiT-B1        & 36.5          & 84.7 & 79.9    & 26.6& 13.8  & 58.7\\
    CFFM \cite{CFFM}              & MiT-B1        & 38.5& 88.6 & 84.1    & -& 15.5  & 29.8          \\
    MRCFA \cite{MRCFA}                & MiT-B1        & 38.9          & 88.8  & 84.4  & -& 16.2 & 40.1    \\
    TV3S  (Ours)                      & MiT-B1        & \textbf{40.0} & \textbf{90.7} & \textbf{87.0} & 36.9& 17.3& 24.7\\ 
    Segformer \cite{xie2021Segformer} & MiT-B2        & 43.9          & 86.0 & 81.2    & 100.8 & 24.8  & 16.2\\
    CFFM \cite{CFFM}              & MiT-B2        & 44.9          & 89.8 & 85.8    & 143.2 & 26.5  & 10.1\\
    MRCFA \cite{MRCFA}                & MiT-B2        & 45.3          & 90.3  & 86.2  & 127.9 & 27.3 & 10.7    \\
    TV3S  (Ours)                      & MiT-B2        & \textbf{46.3} & \textbf{91.5} & \textbf{88.35}& 53.9& 28.3  & 21.9\\ 
    \midrule
    Mask2Former\textsuperscript{\textit{$\dagger$}} \cite{Mask2Former} & R50 & 38.5 & 81.3 & 76.4 & 110.6 & 44.0 & 19.4 \\
    MPVSS \cite{MPVSS} & R50 & 37.5 & 84.1 & 77.2 & 38.9 & 84.1 & 33.9\\
    Mask2Former\textsuperscript{\textit{$\dagger$}} \cite{Mask2Former} & R101 & 39.3 & 82.5 & 77.6 & 141.3 & 63.0 & 16.9\\
    MPVSS \cite{MPVSS} & R101 & 38.8 & 84.8 & 79.6 & 45.1 & 103.1 & 32.3\\
    DeepLabv3+\textsuperscript{\textit{$\dagger$}} \cite{DeepLabv3+} & R101 & 34.7 & 83.2 & 78.2 & 379.0 & 62.7 & 9.2\\
    UperNet\textsuperscript{\textit{$\dagger$}} \cite{UperNet} & R101 & 36.5 & 82.6 & 76.1 & 403.6 & 83.2 & 16.0\\
    PSPNet\textsuperscript{\textit{$\dagger$}} \cite{PSPNet} & R101 & 36.5 & 84.2 & 79.6 & 401.8 & 70.5 & 13.8\\
    OCRNet\textsuperscript{\textit{$\dagger$}} \cite{OCRNet} & R101 & 36.7 & 84.0 & 79.0 & 361.7 & 58.1 & 14.3\\
    TCB\textsuperscript{\textit{$\dagger$}} \cite{VSPW_dataset} & R101 & 37.8 & 87.9 & 84.0 & 1692 & - & -\\
    ETC\textsuperscript{\textit{$\dagger$}} \cite{ETC} & OCRNet & 37.5 & 84.1 & 79.1 & 361.7 & - & - \\
    Segformer \cite{xie2021Segformer} & MiT-B5 & 48.9 & 87.8 & 83.7 & 185.0 & 82.1 & 9.4\\
    CFFM \cite{CFFM} & MiT-B5 & 49.3 & 90.8 & 87.1 & 413.5 & 85.5 & 4.5\\
    MRCFA \cite{MRCFA} & MiT-B5 & \textbf{49.9} & 90.9 & 87.4 & 373.0 & 84.5 & 5.0\\
    TV3S  (Ours) & MiT-B5 & \underline{49.8} & \textbf{91.7} & \textbf{88.7} & 137.0& 85.6 & 14.0\\ 
    \midrule
    Mask2Former\textsuperscript{\textit{$\dagger$}} \cite{Mask2Former} & Swin-T & 41.2 & 84.5 & 80.0 & 114.4 & 47.4 & 17.1\\
    MPVSS \cite{MPVSS} & Swin-T & 39.9 & 85.9 & 80.4 & 39.7 & 114.0 & 32.8\\
    TV3S  (Ours) & Swin-T & \textbf{44.9} & \textbf{88.0} & \textbf{83.5} & 57.3& 31.7 & 22.9\\ 
    Mask2Former\textsuperscript{\textit{$\dagger$}} \cite{Mask2Former} & Swin-S & 42.1 & 84.7 & 79.3 & 152.2 & 68.9 & 14.5\\
    MPVSS \cite{MPVSS} & Swin-S & 40.4 & 86.0 & 80.7 & 47.3 & 108.0 & 30.6\\
    TV3S  (Ours) & Swin-S & \textbf{50.6} & \textbf{89.6} & \textbf{85.8} & 94.1& 53.1 & 19.5\\ 
    \bottomrule
  \end{tabular}}
  \vspace{-6pt}
  \caption{Quantitative comparison of our model with existing methods on the VSPW dataset \cite{VSPW_dataset}. Our model achieves a strong balance between \textit{accuracy}, \textit{model complexity}, and \textit{operational speed}. FPS and FLOPs are calculated with an input resolution of 480 × 853. 
  \footnotesize
  (\textsuperscript{\textit{$\dagger$}}Frame-by-Frame processing)
  }
  \label{tab:VSPW_comparision}
  \vspace{-.1in}
\end{table*}
\section{Experiments}
\label{sec:Experiments}

\subsection{Experimental Setup}
\textbf{Implementation details.}\quad  
The implementation of our approach is based on \texttt{MMSegmentation} codebase and all the experiments including training and inference were conducted with 2 A100 NVIDIA GPUs. The main experiments are done with the backbone same as the SegFormer (Variants with MiT and Swin) which are pre-trained with ImageNet. While the model is aimed to work with any number of frame sequences, during training, the model is trained with just three reference frames, \(\{k_1,k_2,k_3\} = \{-9,-6,-3\}\). More information on training is found in \cref{ssec:Training}. To improve receptive field when processing the features from the backbone, the window size $w$ and the shift $s$ are set to 20 and 10 with number of TV3S blocks $N$ set to 4 (more information on these at \cref{sec:ablation}) and the decoder made with mamba following its default parameters following \cite{mamba_paper} with the input embedding dimension of 256 matching the SegFormer implementation. The images from VSPW dataset \cite{VSPW_dataset} are cropped down to $480\times480$ and are augmented using various augmentation methods during training, including cropping, resizing, flipping and addition of photometric distortion during training. Optimization of the model is done with the use AdamW optimizer and a "poly" learning rate schedule initializing the learning rate at $6e-5$. Testing is performed using the context of full video with the frame receiving the context of all the past frames within the video through the hidden states $\bm{H}$ due to its high efficiency \& effectiveness following \cref{ssec:Inference}. Note that for all cases there was no use of post-processing on the obtained output like in \cite{CRF}. 

\para{Datasets.} The experiments were mainly conducted with the use of VSPW dataset \cite{VSPW_dataset} which stands as one of the largest VSS benchmark. The dataset consists of training, validation and test subsets containing 2,806 clips (198,244 frames), 343 clips (24,502 frames) and 387 clips (28,887 frames) accordingly. The dataset consisting of a rich 124 categories with a dense annotation of frame rate of 15fps contrasts itself from the previously available datasets, which had very sparse annotation with just one frame being annotated every 10s of frames. Furthermore, the dataset covering various different scenarios of both outdoor and indoor scenes makes it suitable for training models to well verify the adaptability of the performance standing as the best benchmark in the field of VSS. While most experiments and training were done with the VSPW dataset \cite{VSPW_dataset}, the proposed method was also evaluated on the Cityscapes dataset \cite{cordts2016cityscapes} which annotates one frame out of every 30 frames for benchmarking the results.

\para{Evaluation metrics.} We use mean IoU (mIoU) and mean video consistency (mVC) as key metrics. The metric mVC evaluates the smoothness of predicted segmentation maps over time, assessing performance in the temporal domain. More formally, given a video clip \(\{\bm{I}_{c}\}^{C_v}_{c=0}\) with ground truth masks $\{\bm{M}_{c}\}^{C_v}_{c=1}$ and predicted outputs $\{\bm{O}_{c}\}^{C_v}_{c=1}$, $VC_n$ is computed as follows:
\begin{equation}
\label{eq:VC}
    \text{VC}_n = {\frac{1}{{C_v}-n+1}} \sum_{i=1}^{{C_v}-n+1} {\frac{(\cap_i^{i+n-1}\bm{M}_i)\cap(\cap_i^{i+n-1}\bm{O}_i)}{\cap_i^{i+n-1}\bm{M}_i}},
\end{equation}
where C\textsubscript{v} $\geq$ n. Once the $VC_n$ of all the videos are computed, their mean is computed to obtain the mVC\textsubscript{n}. \cref{eq:VC} shows that mVC\textsubscript{n} finds the common areas of the predicted masks among frames which indicates the level of consistency of the prediction masks across time. More information on the metric can be found at \cite{VSPW_dataset}.

\setlength\belowcaptionskip{+2ex}
\begin{table}[!t]
  \centering
  \setlength{\tabcolsep}{1.mm}
  \resizebox{\linewidth}{!}{
  \begin{tabular}{l|c|c|c|c|c}
    \toprule
    Methods & Backbones & mIoU$\uparrow$ & GFLOPs$\downarrow$ & Params(M)$\downarrow$ & FPS$\uparrow$ \\
    \midrule
    ETC\textsuperscript{\textit{$\dagger$}} \cite{ETC} & R18 & 71.1 & 434.1 & -- & -- \\
    SegFormer\textsuperscript{\textit{$\dagger$}} \cite{xie2021Segformer} & MiT-B0 & 71.9 & - & 3.7 & 58.5 \\
    CFFM \cite{CFFM} & MiT-B0 & 74.0 & 80.7 & 4.6 & 15.8 \\
    MRCFA \cite{MRCFA} & MiT-B0 & 72.8 & 77.5 & 4.2 & 16.6 \\
    SegFormer\textsuperscript{\textit{$\dagger$}} \cite{xie2021Segformer} & MiT-B1 & 74.1 & - & 13.8 & 46.8 \\
    CFFM \cite{CFFM} & MiT-B1 & 75.1 & 158.7 & 15.4 & 11.7 \\
    MRCFA \cite{MRCFA} & MiT-B1 & 75.1 & 145.0 & 14.9 & 13.0 \\
    TV3S (ours) & MiT-B1 & \textbf{75.6} & 83.6 & 17.3 & 25.1 \\
    \bottomrule
  \end{tabular}}
  \vspace{-5pt}
  \caption{Quantitative comparison of our method with efficient alternative approaches on the Cityscapes dataset \cite{cordts2016cityscapes}, using a resolution of 512 × 1024. 
  \footnotesize
  (\textsuperscript{\textit{$\dagger$}}Frame-by-Frame processing)
  }
  \label{tab:City_comparision}
  \vspace{-.1in}
\end{table}

\begin{table}[!t]
    \centering
    \renewcommand{\arraystretch}{1.} 
    \setlength{\tabcolsep}{5mm}
    \resizebox{\linewidth}{!}{
    \begin{tabular}{c|c|c|c}
        \toprule
        \multirow{2}{*}{Models} & \multicolumn{3}{c}{Evaluation (mIoU)} \\ \cmidrule{2-4} 
        & Add & Concat & Direct \\ 
        \midrule
        1 TSS (No Shift) & \multicolumn{1}{c|}{37.6} & \multicolumn{1}{c|}{37.8}          & 38.0\\ 
        TV3S (No Shift)  & \multicolumn{1}{c|}{38.1} & \multicolumn{1}{c|}{38.4} & 38.9\\ 
        TV3S (Shift) & \multicolumn{1}{c|}{\textbf{38.5}} & \multicolumn{1}{c|}{\textbf{39.3}} & \textbf{39.5}\\ 
        \bottomrule
    \end{tabular}}
    \vspace{-5pt}
    \caption{Evaluation on the implication of the shifted mechanism and the output method.}
    \label{tab:Eval_internal}
    \vspace{-.1in}
\end{table}

\setlength\belowcaptionskip{+0.5ex}
\begin{table}[!t]
    \centering
    \setlength{\tabcolsep}{5mm}
    \resizebox{.95\linewidth}{!}{
    \begin{tabular}{c|c|c|c} \toprule
        Window sizes & mIoU  & mVC\textsubscript{8} & mVC\textsubscript{16} \\ \midrule
        4      & 38.5& 89.0  & 84.6   \\
        6      & 38.6& 89.2  & 84.7   \\
        12     & 39.3& 89.3 & 85.0   \\
        16     & 39.2& 89.3 & 85.0   \\
        20     & \textbf{40.0}& \textbf{90.7} & \textbf{87.0} \\
        28     & 40.0& 88.8 & 84.2  \\     
        36     & 39.9& 89.1 & 84.9  \\     
        \bottomrule
    \end{tabular}}
    \vspace{-.1in}
    \caption{Impact of the window size on model performance.}
    \label{tab:Eval_window}
    \vspace{-.13in}
\end{table}

\setlength\belowcaptionskip{+1ex}
\begin{table}[!t]
    \centering
    \setlength{\tabcolsep}{3.3mm}
    \resizebox{\linewidth}{!}{
    \begin{tabular}{c|c|c|c|c} \toprule
        Backbones & TV3S Blocks & mIoU & mVC\textsubscript{8} & mVC\textsubscript{16} \\ 
        \midrule
        \multirow{4}{*}{MiT-B1} & 
                  1      & 36.2& 88.1 & 83.6  \\
                & 2      & 37.4& 88.6 & 84.3  \\
                & 3      & 37.6& 88.5 & 83.5  \\
                & 4      & \textbf{40.0}& \textbf{90.7} & \textbf{87.0}  \\
        \midrule
    \end{tabular}}
    \vspace{-.1in}
    \caption{Performance metrics based on the number of TV3S blocks in the model.}
    \label{tab:Eval_train_Block}
    \vspace{-.05in}
\end{table}

\subsection{Comparison with State-of-the-art Models}
\begin{figure*}[!t]
    \centering
    \includegraphics[width=0.96\linewidth]{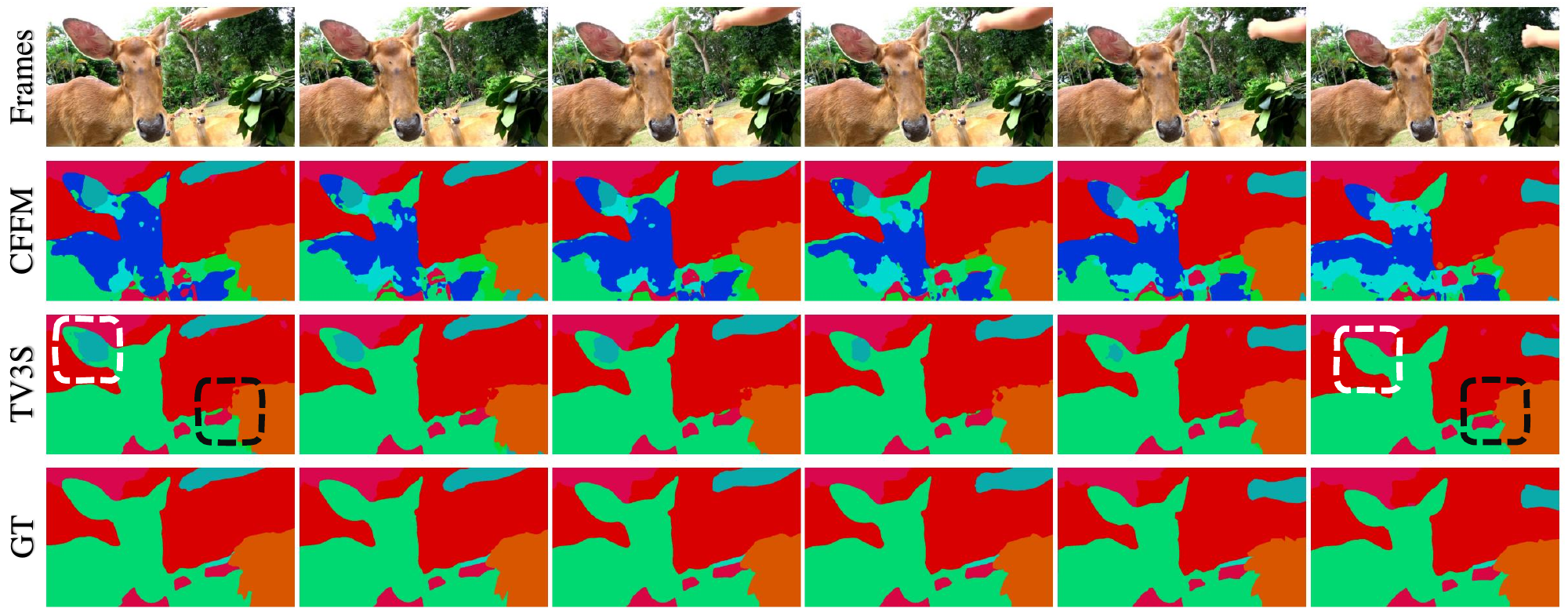}
    \caption{Qualitative example of our TV3S architecture compared with the current baseline. This displays both improved performance in performing spatial predictions and utilizing the temporal information to produce temporally consistent segmentation results.}
    \label{fig:vis_comp}
    \vspace{-.15in}
\end{figure*}

Our model is compared against state-of-the-art models on the VSPW dataset \cite{VSPW_dataset}, as shown in \cref{tab:VSPW_comparision}. The table is divided into three groups based on model size and backbone, offering insights at varying scales. In the first group with small models ($<$30M parameters), our method outperforms the baselines, demonstrating efficiency even with limited model capacity. In the second group with larger models ($>$30M parameters), TV3S achieves near-state-of-the-art performance by effectively capturing rich contextual information. The third group focuses on the Swin Transformer backbone, where our model excels with over 8 mIoU ahead of the next best, highlighting its ability to preserve temporal correlations and ensure spatial accuracy. Overall, our method demonstrates superior visual consistency across all groups, further illustrated by \cref{fig:vis_comp} showcasing temporally consistent segmentation.

We also present results on the Cityscapes dataset \cite{cordts2016cityscapes} in \cref{tab:City_comparision}, using smaller model variants at an input resolution of \(1024 \times 512\). Due to the dataset’s annotation structure, metrics such as mVC\textsubscript{8} and mVC\textsubscript{16} are not applicable, but our model still achieves top performance in mIoU, underscoring its strong generalizability across datasets.

\setlength\belowcaptionskip{-0.5ex}

\subsection{Ablation Study}
\label{sec:ablation}
All ablation studies were conducted on the VSPW dataset \cite{VSPW_dataset} using the MiT-B1 backbone and following the same training and inference strategies as previously described.

\para{Impact of shifted representations and feature integration.}
We evaluated the effectiveness of different TSS/TV3S decoder (see \cref{tab:Eval_internal}), including one TSS layer, unshifted TV3S, and shifted TV3S. Despite the only difference between unshifted and shifted TV3S being the feature representation, the results clearly show that shifted representation significantly outperforms its unshifted counterparts, highlighting their effectiveness. We further tested adding feature maps through addition and concatenation before the final prediction, but both methods reduced performance, suggesting that reintroducing spatial information after temporal modeling disrupts temporal coherence.
\begin{table}[!t]
    \centering
    \setlength{\tabcolsep}{4.5mm}
    \resizebox{\linewidth}{!}{
    \begin{tabular}{c|c|c|c} \toprule
        Number of Frames & mIoU & mVC\textsubscript{8} & mVC\textsubscript{16} \\ 
        \midrule
        1      & 37.9& 84.5 & 79.1     \\
        2      & 39.2& 86.9 & 81.8     \\
        4      & 39.5& 88.9 & 84.2     \\
        8      & 39.7& 90.4 & 85.9     \\
        16     & 39.7& 91.2 & 87.2     \\
        32     & \textbf{39.7}& \textbf{91.5} & \textbf{87.8}     \\ 
        \bottomrule
    \end{tabular}}
    \vspace{-.1in}
    \caption{Performance based on the amount of temporal information available during inference}
    \label{tab:Eval_inferTemporal}
\end{table}

\para{Impact of window size.} 
TV3S establishes temporal relationships with window size controlling the spatial context during processing. Larger windows help summarize stable scenes but may overload the model in dynamic settings, while smaller windows enhance understanding in fast-changing environments but may miss correlations in slower scenes. As shown in \cref{tab:Eval_window}, the optimal performance is achieved with a $20\times 20$ window size, striking a balance between spatial context and model efficiency. 

\para{Effect on the number of TV3S blocks.} 
In our experiments, we assessed the impact of varying the number of TV3S blocks, as shown in \cref{tab:Eval_train_Block}. Increasing the number of blocks improved mIoU from 36.18 with one block to 40.00 with four blocks, along with enhanced temporal consistency, as shown by the mVC\textsubscript{8} and mVC\textsubscript{16} scores. This upward trend suggests that stacking TV3S blocks better aggregates temporal features, capturing complex relationships without introducing excessive redundancy.

\subsection{Temporal Context}
\textbf{Importance of long temporal context.}\quad 
  Given Mamba's RNN-like structure, we conducted an evaluation of its performance on the VSPW dataset \cite{VSPW_dataset} with varying temporal context sizes that ranged from 1 to 32 frames. In our approach, the videos that were shorter than 32 frames were excluded from the and at the same time the first 32 frames were omitted during inference to ensure a fair and unbiased evaluation of the model's capabilities. The findings presented in \cref{tab:Eval_inferTemporal} indicate that there is a notable improvement in performance as the temporal context sizes increase. However, this enhancement reaches a plateau once the temporal information saturates. This saturation effect is likely attributed to the limited changes that can be captured within smaller temporal windows, which causes diminishing returns in performance improvements beyond a certain threshold.

\section{Conclusion}

The work proposes a TV3S architecture that addresses the challenges of VSS by capturing temporal dynamics. The proposed structure is designed to achieve good computational efficiency and scalability. By leveraging independent processing of spatial blocks through state space models enhanced with Mamba \cite{mamba_paper}, our approach enables parallel computation during both training and inference. This design mitigates the time delay and high memory demand typically associated with sequential processing in traditional state space and recurrent models, making it highly suitable for long, high-resolution video sequences. Extensive experiments on the VSPW \cite{VSPW_dataset} and Cityscapes \cite{cordts2016cityscapes} datasets show that the TV3S architecture not only surpasses existing methods in segmentation accuracy but also significantly improves temporal prediction consistency.

\para{Future works.}
Our model enhances the encoder's focus on key temporal features, integrating spatial and temporal information for improved video frame segmentation. Its adaptability extends to tasks like object detection and action recognition, while its multi-modal data integration offers new research opportunities in audio-visual learning, emphasizing temporal synchronization.
\section*{Acknowledgment}
This work is supported by the Ministry of Education Singapore Tier 1 project funding grant No. MOE Tier 1 RG98/24 and by Agency for Science, Technology and Research (A*STAR) under its MTC Programmatic Funds (Grant No. M23L7b0021).

{
    \small
    \bibliographystyle{ieeenat_fullname}
    \bibliography{main}

\begin{thebibliography}{75}
\providecommand{\natexlab}[1]{#1}
\providecommand{\url}[1]{\texttt{#1}}
\expandafter\ifx\csname urlstyle\endcsname\relax
  \providecommand{\doi}[1]{doi: #1}\else
  \providecommand{\doi}{doi: \begingroup \urlstyle{rm}\Url}\fi

\bibitem[An et~al.(2023)An, Sun, Wu, Tang, and Van~Gool]{an2023temporal}
Zhaochong An, Guolei Sun, Zongwei Wu, Hao Tang, and Luc Van~Gool.
\newblock Temporal-aware hierarchical mask classification for video semantic segmentation.
\newblock In \emph{BMVC}, 2023.

\bibitem[Atluri et~al.(2018)Atluri, Karpatne, and Kumar]{atluri2018spatio}
Gowtham Atluri, Anuj Karpatne, and Vipin Kumar.
\newblock Spatio-temporal data mining: A survey of problems and methods.
\newblock \emph{ACM Computing Surveys}, 51\penalty0 (4):\penalty0 1--41, 2018.

\bibitem[Ballas et~al.(2016)Ballas, Yao, Pal, and Courville]{ballas2016delving}
Nicolas Ballas, Li Yao, Chris Pal, and Aaron Courville.
\newblock Delving deeper into convolutional networks for learning video representations.
\newblock In \emph{ICLR}, 2016.

\bibitem[Carion et~al.(2020)Carion, Massa, Synnaeve, Usunier, Kirillov, and Zagoruyko]{DETR}
Nicolas Carion, Francisco Massa, Gabriel Synnaeve, Nicolas Usunier, Alexander Kirillov, and Sergey Zagoruyko.
\newblock End-to-end object detection with transformers.
\newblock In \emph{ECCV}, pages 213--229, 2020.

\bibitem[Chen et~al.(2017)Chen, Papandreou, Kokkinos, Murphy, and Yuille]{DeepLab}
Liang-Chieh Chen, George Papandreou, Iasonas Kokkinos, Kevin Murphy, and Alan~L Yuille.
\newblock {DeepLab}: Semantic image segmentation with deep convolutional nets, atrous convolution, and fully connected {CRFs}.
\newblock \emph{IEEE TPAMI}, 40\penalty0 (4):\penalty0 834--848, 2017.

\bibitem[Chen et~al.(2018)Chen, Zhu, Papandreou, Schroff, and Adam]{DeepLabv3+}
Liang-Chieh Chen, Yukun Zhu, George Papandreou, Florian Schroff, and Hartwig Adam.
\newblock Encoder-decoder with atrous separable convolution for semantic image segmentation.
\newblock In \emph{ECCV}, pages 801--818, 2018.

\bibitem[Cheng et~al.(2022)Cheng, Misra, Schwing, Kirillov, and Girdhar]{Mask2Former}
Bowen Cheng, Ishan Misra, Alexander~G Schwing, Alexander Kirillov, and Rohit Girdhar.
\newblock Masked-attention mask transformer for universal image segmentation.
\newblock In \emph{CVPR}, pages 1290--1299, 2022.

\bibitem[Cheng et~al.(2017)Cheng, Tsai, Wang, and Yang]{cheng2017segflow}
Jingchun Cheng, Yi-Hsuan Tsai, Shengjin Wang, and Ming-Hsuan Yang.
\newblock {SegFlow}: Joint learning for video object segmentation and optical flow.
\newblock In \emph{ICCV}, pages 686--695, 2017.

\bibitem[Cordts et~al.(2016)Cordts, Omran, Ramos, Rehfeld, Enzweiler, Benenson, Franke, Roth, and Schiele]{cordts2016cityscapes}
Marius Cordts, Mohamed Omran, Sebastian Ramos, Timo Rehfeld, Markus Enzweiler, Rodrigo Benenson, Uwe Franke, Stefan Roth, and Bernt Schiele.
\newblock The {C}ityscapes dataset for semantic urban scene understanding.
\newblock In \emph{CVPR}, pages 3213--3223, 2016.

\bibitem[Ding et~al.(2018)Ding, Jiang, Shuai, Liu, and Wang]{CCL}
Henghui Ding, Xudong Jiang, Bing Shuai, Ai~Qun Liu, and Gang Wang.
\newblock Context contrasted feature and gated multi-scale aggregation for scene segmentation.
\newblock In \emph{CVPR}, pages 2393--2402, 2018.

\bibitem[Ding et~al.(2023{\natexlab{a}})Ding, Liu, He, Jiang, and Loy]{MeViS}
Henghui Ding, Chang Liu, Shuting He, Xudong Jiang, and Chen~Change Loy.
\newblock {MeViS}: {A} large-scale benchmark for video segmentation with motion expressions.
\newblock In \emph{ICCV}, pages 2694--2703, 2023{\natexlab{a}}.

\bibitem[Ding et~al.(2023{\natexlab{b}})Ding, Liu, He, Jiang, Torr, and Bai]{MOSE}
Henghui Ding, Chang Liu, Shuting He, Xudong Jiang, Philip~HS Torr, and Song Bai.
\newblock {MOSE}: A new dataset for video object segmentation in complex scenes.
\newblock In \emph{ICCV}, pages 20224--20234, 2023{\natexlab{b}}.

\bibitem[Ding et~al.(2020)Ding, Wang, Zhou, Shi, Lu, and Luo]{ding2020every}
Mingyu Ding, Zhe Wang, Bolei Zhou, Jianping Shi, Zhiwu Lu, and Ping Luo.
\newblock Every frame counts: Joint learning of video segmentation and optical flow.
\newblock In \emph{AAAI}, pages 10713--10720, 2020.

\bibitem[Dosovitskiy(2021)]{ViT}
Alexey Dosovitskiy.
\newblock An image is worth 16x16 words: Transformers for image recognition at scale.
\newblock In \emph{ICLR}, 2021.

\bibitem[Emre~Yurdakul and Yemez(2017)]{emre2017semantic}
Ekrem Emre~Yurdakul and Yucel Yemez.
\newblock Semantic segmentation of rgbd videos with recurrent fully convolutional neural networks.
\newblock In \emph{ICCV workshops}, pages 367--374, 2017.

\bibitem[Fu et~al.(2019)Fu, Liu, Tian, Li, Bao, Fang, and Lu]{fu2019dual}
Jun Fu, Jing Liu, Haijie Tian, Yong Li, Yongjun Bao, Zhiwei Fang, and Hanqing Lu.
\newblock Dual attention network for scene segmentation.
\newblock In \emph{CVPR}, pages 3146--3154, 2019.

\bibitem[Gao et~al.(2023)Gao, Wang, Zhuang, Zhang, and Li]{gao2023exploit}
Yuan Gao, Zilei Wang, Jiafan Zhuang, Yixin Zhang, and Junjie Li.
\newblock Exploit domain-robust optical flow in domain adaptive video semantic segmentation.
\newblock In \emph{AAAI}, pages 641--649, 2023.

\bibitem[Geiger et~al.(2013)Geiger, Lenz, Stiller, and Urtasun]{geiger2013vision}
Andreas Geiger, Philip Lenz, Christoph Stiller, and Raquel Urtasun.
\newblock Vision meets robotics: The {KITTI} dataset.
\newblock \emph{The International Journal of Robotics Research}, 32\penalty0 (11):\penalty0 1231--1237, 2013.

\bibitem[Gu and Dao(2023)]{mamba_paper}
Albert Gu and Tri Dao.
\newblock Mamba: Linear-time sequence modeling with selective state spaces.
\newblock \emph{arXiv preprint arXiv:2312.00752}, 2023.

\bibitem[Gu et~al.(2020)Gu, Dao, Ermon, Rudra, and R{\'e}]{gu2020hippo}
Albert Gu, Tri Dao, Stefano Ermon, Atri Rudra, and Christopher R{\'e}.
\newblock {HiPPO}: Recurrent memory with optimal polynomial projections.
\newblock In \emph{NeurIPS}, pages 1474--1487, 2020.

\bibitem[Gu et~al.(2021{\natexlab{a}})Gu, Goel, and R{\'e}]{gu2021efficiently}
Albert Gu, Karan Goel, and Christopher R{\'e}.
\newblock Efficiently modeling long sequences with structured state spaces.
\newblock In \emph{ICLR}, 2021{\natexlab{a}}.

\bibitem[Gu et~al.(2021{\natexlab{b}})Gu, Johnson, Goel, Saab, Dao, Rudra, and R{\'e}]{gu2021combining}
Albert Gu, Isys Johnson, Karan Goel, Khaled Saab, Tri Dao, Atri Rudra, and Christopher R{\'e}.
\newblock Combining recurrent, convolutional, and continuous-time models with linear state space layers.
\newblock In \emph{NeurIPS}, pages 572--585, 2021{\natexlab{b}}.

\bibitem[Guo et~al.(2018)Guo, Jin, Gotz, Du, Zha, and Cao]{guo2018visual}
Shunan Guo, Zhuochen Jin, David Gotz, Fan Du, Hongyuan Zha, and Nan Cao.
\newblock Visual progression analysis of event sequence data.
\newblock \emph{IEEE TVCG}, 25\penalty0 (1):\penalty0 417--426, 2018.

\bibitem[Hu et~al.(2018)Hu, Shen, and Sun]{hu2018squeeze}
Jie Hu, Li Shen, and Gang Sun.
\newblock Squeeze-and-excitation networks.
\newblock In \emph{CVPR}, pages 7132--7141, 2018.

\bibitem[Hu et~al.(2020)Hu, Caba, Wang, Lin, Sclaroff, and Perazzi]{hu2020temporally}
Ping Hu, Fabian Caba, Oliver Wang, Zhe Lin, Stan Sclaroff, and Federico Perazzi.
\newblock Temporally distributed networks for fast video semantic segmentation.
\newblock In \emph{CVPR}, pages 8818--8827, 2020.

\bibitem[Jain et~al.(2019)Jain, Wang, and Gonzalez]{Accel}
Samvit Jain, Xin Wang, and Joseph~E Gonzalez.
\newblock Accel: A corrective fusion network for efficient semantic segmentation on video.
\newblock In \emph{CVPR}, pages 8866--8875, 2019.

\bibitem[Kr{\"a}henb{\"u}hl and Koltun(2011)]{CRF}
Philipp Kr{\"a}henb{\"u}hl and Vladlen Koltun.
\newblock Efficient inference in fully connected {CRFs} with gaussian edge potentials.
\newblock In \emph{NeurIPS}, pages 109--117, 2011.

\bibitem[Kundu et~al.(2016)Kundu, Vineet, and Koltun]{kundu2016feature}
Abhijit Kundu, Vibhav Vineet, and Vladlen Koltun.
\newblock Feature space optimization for semantic video segmentation.
\newblock In \emph{CVPR}, pages 3168--3175, 2016.

\bibitem[Lao et~al.(2023)Lao, Hong, Guo, Zhang, Wang, Chen, and Chu]{lao2023simultaneously}
Jiangwei Lao, Weixiang Hong, Xin Guo, Yingying Zhang, Jian Wang, Jingdong Chen, and Wei Chu.
\newblock Simultaneously short-and long-term temporal modeling for semi-supervised video semantic segmentation.
\newblock In \emph{CVPR}, pages 14763--14772, 2023.

\bibitem[Li et~al.(2019)Li, Zhao, Fu, Wu, and Liu]{li2019attention}
Jiangyun Li, Yikai Zhao, Jun Fu, Jiajia Wu, and Jing Liu.
\newblock Attention-guided network for semantic video segmentation.
\newblock \emph{IEEE Access}, 7:\penalty0 140680--140689, 2019.

\bibitem[Li et~al.(2021)Li, Wang, Chen, Niu, Si, Qian, and Zhang]{li2021video}
Jiangtong Li, Wentao Wang, Junjie Chen, Li Niu, Jianlou Si, Chen Qian, and Liqing Zhang.
\newblock Video semantic segmentation via sparse temporal transformer.
\newblock In \emph{ACM MM}, pages 59--68, 2021.

\bibitem[Li et~al.(2025)Li, Li, Wang, He, Wang, Wang, and Qiao]{li2025videomamba}
Kunchang Li, Xinhao Li, Yi Wang, Yinan He, Yali Wang, Limin Wang, and Yu Qiao.
\newblock Video{M}amba: State space model for efficient video understanding.
\newblock In \emph{ECCV}, pages 237--255, 2025.

\bibitem[Li et~al.(2022)Li, Zhang, Pang, Chen, Cheng, Tong, and Loy]{li2022video}
Xiangtai Li, Wenwei Zhang, Jiangmiao Pang, Kai Chen, Guangliang Cheng, Yunhai Tong, and Chen~Change Loy.
\newblock Video {K-Net}: A simple, strong, and unified baseline for video segmentation.
\newblock In \emph{CVPR}, pages 18847--18857, 2022.

\bibitem[Li et~al.(2023)Li, Yuan, Zhang, Cheng, Pang, and Loy]{li2023tube}
Xiangtai Li, Haobo Yuan, Wenwei Zhang, Guangliang Cheng, Jiangmiao Pang, and Chen~Change Loy.
\newblock Tube-{Link}: A flexible cross tube framework for universal video segmentation.
\newblock In \emph{ICCV}, pages 13923--13933, 2023.

\bibitem[Li et~al.(2018)Li, Shi, and Lin]{li2018low}
Yule Li, Jianping Shi, and Dahua Lin.
\newblock Low-latency video semantic segmentation.
\newblock In \emph{CVPR}, pages 5997--6005, 2018.

\bibitem[Litjens et~al.(2017)Litjens, Kooi, Bejnordi, Setio, Ciompi, Ghafoorian, Van Der~Laak, Van~Ginneken, and S{\'a}nchez]{litjens2017survey}
Geert Litjens, Thijs Kooi, Babak~Ehteshami Bejnordi, Arnaud Arindra~Adiyoso Setio, Francesco Ciompi, Mohsen Ghafoorian, Jeroen~Awm Van Der~Laak, Bram Van~Ginneken, and Clara~I S{\'a}nchez.
\newblock A survey on deep learning in medical image analysis.
\newblock \emph{Medical Image Analysis}, 42:\penalty0 60--88, 2017.

\bibitem[Liu et~al.(2020)Liu, Shen, Yu, and Wang]{ETC}
Yifan Liu, Chunhua Shen, Changqian Yu, and Jingdong Wang.
\newblock Efficient semantic video segmentation with per-frame inference.
\newblock In \emph{ECCV}, pages 352--368, 2020.

\bibitem[Liu et~al.(2024)Liu, Tian, Zhao, Yu, Xie, Wang, Ye, and Liu]{liu2024vmambavisualstatespace}
Yue Liu, Yunjie Tian, Yuzhong Zhao, Hongtian Yu, Lingxi Xie, Yaowei Wang, Qixiang Ye, and Yunfan Liu.
\newblock {VMamba}: Visual state space model.
\newblock \emph{arXiv preprint arXiv:2401.10166}, 2024.

\bibitem[Liu et~al.(2021)Liu, Lin, Cao, Hu, Wei, Zhang, Lin, and Guo]{Swin}
Ze Liu, Yutong Lin, Yue Cao, Han Hu, Yixuan Wei, Zheng Zhang, Stephen Lin, and Baining Guo.
\newblock Swin transformer: Hierarchical vision transformer using shifted windows.
\newblock In \emph{ICCV}, pages 10012--10022, 2021.

\bibitem[Long et~al.(2015)Long, Shelhamer, and Darrell]{FCN}
Jonathan Long, Evan Shelhamer, and Trevor Darrell.
\newblock Fully convolutional networks for semantic segmentation.
\newblock In \emph{CVPR}, pages 3431--3440, 2015.

\bibitem[Ma et~al.(2024)Ma, Li, and Wang]{ma2024u}
Jun Ma, Feifei Li, and Bo Wang.
\newblock {U-Mamba}: Enhancing long-range dependency for biomedical image segmentation.
\newblock \emph{arXiv preprint arXiv:2401.04722}, 2024.

\bibitem[Miao et~al.(2021)Miao, Wei, Wu, Liang, Li, and Yang]{VSPW_dataset}
Jiaxu Miao, Yunchao Wei, Yu Wu, Chen Liang, Guangrui Li, and Yi Yang.
\newblock {VSPW}: A large-scale dataset for video scene parsing in the wild.
\newblock In \emph{CVPR}, pages 4133--4143, 2021.

\bibitem[Nabavi et~al.(2018)Nabavi, Rochan, and Wang]{nabavi2018future}
Seyed~Shahabeddin Nabavi, Mrigank Rochan, and Yang Wang.
\newblock Future semantic segmentation with convolutional lstm.
\newblock \emph{CoRR}, 2018.

\bibitem[Nilsson and Sminchisescu(2018)]{nilsson2018semantic}
David Nilsson and Cristian Sminchisescu.
\newblock Semantic video segmentation by gated recurrent flow propagation.
\newblock In \emph{CVPR}, pages 6819--6828, 2018.

\bibitem[O~Pinheiro et~al.(2015)O~Pinheiro, Collobert, and Doll{\'a}r]{o2015learning}
Pedro~O O~Pinheiro, Ronan Collobert, and Piotr Doll{\'a}r.
\newblock Learning to segment object candidates.
\newblock In \emph{NeurIPS}, pages 1441--1447, 2015.

\bibitem[Pfeuffer et~al.(2019)Pfeuffer, Schulz, and Dietmayer]{pfeuffer2019semantic}
Andreas Pfeuffer, Karina Schulz, and Klaus Dietmayer.
\newblock Semantic segmentation of video sequences with convolutional {LSTMs}.
\newblock In \emph{IEEE Intelligent Vehicles Symposium}, pages 1441--1447, 2019.

\bibitem[Qian et~al.(2024)Qian, Lin, See, and Li]{qian2024controllable}
Rui Qian, Weiyao Lin, John See, and Dian Li.
\newblock Controllable augmentations for video representation learning.
\newblock \emph{Visual Intelligence}, 2\penalty0 (1):\penalty0 1, 2024.

\bibitem[Ronneberger et~al.(2015)Ronneberger, Fischer, and Brox]{ronneberger2015u}
Olaf Ronneberger, Philipp Fischer, and Thomas Brox.
\newblock {U-Net}: Convolutional networks for biomedical image segmentation.
\newblock In \emph{International Conference on Medical Image Computing and Computer-Assisted Intervention}, pages 234--241, 2015.

\bibitem[Sandler et~al.(2018)Sandler, Howard, Zhu, Zhmoginov, and Chen]{mobilenetv2}
Mark Sandler, Andrew Howard, Menglong Zhu, Andrey Zhmoginov, and Liang-Chieh Chen.
\newblock {MobileNetV2}: Inverted residuals and linear bottlenecks.
\newblock In \emph{CVPR}, pages 4510--4520, 2018.

\bibitem[Sevilla-Lara et~al.(2016)Sevilla-Lara, Sun, Jampani, and Black]{sevilla2016optical}
Laura Sevilla-Lara, Deqing Sun, Varun Jampani, and Michael~J Black.
\newblock Optical flow with semantic segmentation and localized layers.
\newblock In \emph{CVPR}, pages 3889--3898, 2016.

\bibitem[Shelhamer et~al.(2016)Shelhamer, Rakelly, Hoffman, and Darrell]{shelhamer2016clockwork}
Evan Shelhamer, Kate Rakelly, Judy Hoffman, and Trevor Darrell.
\newblock Clockwork convnets for video semantic segmentation.
\newblock In \emph{ECCV}, pages 852--868, 2016.

\bibitem[Shi et~al.(2015)Shi, Chen, Wang, Yeung, Wong, and Woo]{shi2015convolutional}
Xingjian Shi, Zhourong Chen, Hao Wang, Dit-Yan Yeung, Wai-Kin Wong, and Wang-chun Woo.
\newblock Convolutional {LSTM} network: A machine learning approach for precipitation nowcasting.
\newblock In \emph{NeurIPS}, pages 802--810, 2015.

\bibitem[Siam et~al.(2018)Siam, Gamal, Abdel-Razek, Yogamani, Jagersand, and Zhang]{siam2018comparative}
Mennatullah Siam, Mostafa Gamal, Moemen Abdel-Razek, Senthil Yogamani, Martin Jagersand, and Hong Zhang.
\newblock A comparative study of real-time semantic segmentation for autonomous driving.
\newblock In \emph{CVPR Workshop}, pages 587--597, 2018.

\bibitem[Siam et~al.(2021)Siam, Kendall, and Jagersand]{siam2021video}
Mennatullah Siam, Alex Kendall, and Martin Jagersand.
\newblock Video class agnostic segmentation benchmark for autonomous driving.
\newblock In \emph{CVPR}, pages 2825--2834, 2021.

\bibitem[Srivastava et~al.(2015)Srivastava, Mansimov, and Salakhudinov]{srivastava2015unsupervised}
Nitish Srivastava, Elman Mansimov, and Ruslan Salakhudinov.
\newblock Unsupervised learning of video representations using lstms.
\newblock In \emph{ICML}, pages 843--852, 2015.

\bibitem[Strudel et~al.(2021)Strudel, Garcia, Laptev, and Schmid]{Segmenter}
Robin Strudel, Ricardo Garcia, Ivan Laptev, and Cordelia Schmid.
\newblock Segmenter: Transformer for semantic segmentation.
\newblock In \emph{ICCV}, pages 7262--7272, 2021.

\bibitem[Sun et~al.(2022{\natexlab{a}})Sun, Liu, Ding, Probst, and Van~Gool]{CFFM}
Guolei Sun, Yun Liu, Henghui Ding, Thomas Probst, and Luc Van~Gool.
\newblock Coarse-to-fine feature mining for video semantic segmentation.
\newblock In \emph{CVPR}, pages 3126--3137, 2022{\natexlab{a}}.

\bibitem[Sun et~al.(2022{\natexlab{b}})Sun, Liu, Tang, Chhatkuli, Zhang, and Van~Gool]{MRCFA}
Guolei Sun, Yun Liu, Hao Tang, Ajad Chhatkuli, Le Zhang, and Luc Van~Gool.
\newblock Mining relations among cross-frame affinities for video semantic segmentation.
\newblock In \emph{ECCV}, pages 522--539, 2022{\natexlab{b}}.

\bibitem[Sun et~al.(2024)Sun, Liu, Ding, Wu, and Van~Gool]{CFFM++}
Guolei Sun, Yun Liu, Henghui Ding, Min Wu, and Luc Van~Gool.
\newblock Learning local and global temporal contexts for video semantic segmentation.
\newblock \emph{IEEE TPAMI}, 46\penalty0 (10):\penalty0 6919--6934, 2024.

\bibitem[Tang et~al.(2024)Tang, Dong, Tang, Chu, and Liang]{tang2024vmrnn}
Yujin Tang, Peijie Dong, Zhenheng Tang, Xiaowen Chu, and Junwei Liang.
\newblock {VMRNN}: Integrating vision mamba and {LSTM} for efficient and accurate spatiotemporal forecasting.
\newblock In \emph{CVPR Workshop}, pages 5663--5673, 2024.

\bibitem[Tsakanikas and Dagiuklas(2018)]{tsakanikas2018video}
Vassilios Tsakanikas and Tasos Dagiuklas.
\newblock Video surveillance systems-current status and future trends.
\newblock \emph{Computers \& Electrical Engineering}, 70:\penalty0 736--753, 2018.

\bibitem[Valipour et~al.(2017)Valipour, Siam, Jagersand, and Ray]{valipour2017recurrent}
Sepehr Valipour, Mennatullah Siam, Martin Jagersand, and Nilanjan Ray.
\newblock Recurrent fully convolutional networks for video segmentation.
\newblock In \emph{IEEE Winter Conference on Applications of Computer Vision}, pages 29--36, 2017.

\bibitem[Vaswani et~al.(2017)Vaswani, Shazeer, Parmar, Uszkoreit, Jones, Gomez, Kaiser, and Polosukhin]{vaswani2017attention}
Ashish Vaswani, Noam Shazeer, Niki Parmar, Jakob Uszkoreit, Llion Jones, Aidan~N Gomez, \L~ukasz Kaiser, and Illia Polosukhin.
\newblock Attention is all you need.
\newblock In \emph{NeurIPS}, pages 5998--6008, 2017.

\bibitem[Wang et~al.(2021{\natexlab{a}})Wang, Li, Nakashima, Kawasaki, Nagahara, and Yagi]{wang2021noisy}
Bowen Wang, Liangzhi Li, Yuta Nakashima, Ryo Kawasaki, Hajime Nagahara, and Yasushi Yagi.
\newblock Noisy-{LSTM}: Improving temporal awareness for video semantic segmentation.
\newblock \emph{IEEE Access}, 9:\penalty0 46810--46820, 2021{\natexlab{a}}.

\bibitem[Wang et~al.(2021{\natexlab{b}})Wang, Wang, and Liu]{wang2021temporal}
Hao Wang, Weining Wang, and Jing Liu.
\newblock Temporal memory attention for video semantic segmentation.
\newblock In \emph{ICIP}, pages 2254--2258, 2021{\natexlab{b}}.

\bibitem[Wang et~al.(2018)Wang, Girshick, Gupta, and He]{wang2018non}
Xiaolong Wang, Ross Girshick, Abhinav Gupta, and Kaiming He.
\newblock Non-local neural networks.
\newblock In \emph{CVPR}, pages 7794--7803, 2018.

\bibitem[Weng et~al.(2024)Weng, Han, He, Li, Yao, Chang, and Zhuang]{MPVSS}
Yuetian Weng, Mingfei Han, Haoyu He, Mingjie Li, Lina Yao, Xiaojun Chang, and Bohan Zhuang.
\newblock Mask propagation for efficient video semantic segmentation.
\newblock In \emph{NeurIPS}, pages 7170--7183, 2024.

\bibitem[Xiao et~al.(2018)Xiao, Liu, Zhou, Jiang, and Sun]{UperNet}
Tete Xiao, Yingcheng Liu, Bolei Zhou, Yuning Jiang, and Jian Sun.
\newblock Unified perceptual parsing for scene understanding.
\newblock In \emph{ECCV}, pages 418--434, 2018.

\bibitem[Xie et~al.(2021)Xie, Wang, Yu, Anandkumar, Alvarez, and Luo]{xie2021Segformer}
Enze Xie, Wenhai Wang, Zhiding Yu, Anima Anandkumar, Jose~M Alvarez, and Ping Luo.
\newblock {SegFormer}: Simple and efficient design for semantic segmentation with transformers.
\newblock In \emph{NeurIPS}, pages 12077--12090, 2021.

\bibitem[Xu et~al.(2012)Xu, Xiong, and Corso]{xu2012streaming}
Chenliang Xu, Caiming Xiong, and Jason~J Corso.
\newblock Streaming hierarchical video segmentation.
\newblock In \emph{ECCV}, pages 626--639, 2012.

\bibitem[Yuan et~al.(2020)Yuan, Chen, and Wang]{OCRNet}
Yuhui Yuan, Xilin Chen, and Jingdong Wang.
\newblock Object-contextual representations for semantic segmentation.
\newblock In \emph{ECCV}, pages 173--190, 2020.

\bibitem[Zhao et~al.(2017)Zhao, Shi, Qi, Wang, and Jia]{PSPNet}
Hengshuang Zhao, Jianping Shi, Xiaojuan Qi, Xiaogang Wang, and Jiaya Jia.
\newblock Pyramid scene parsing network.
\newblock In \emph{CVPR}, pages 2881--2890, 2017.

\bibitem[Zheng et~al.(2021)Zheng, Lu, Zhao, Zhu, Luo, Wang, Fu, Feng, Xiang, Torr, et~al.]{SETR}
Sixiao Zheng, Jiachen Lu, Hengshuang Zhao, Xiatian Zhu, Zekun Luo, Yabiao Wang, Yanwei Fu, Jianfeng Feng, Tao Xiang, Philip~HS Torr, et~al.
\newblock Rethinking semantic segmentation from a sequence-to-sequence perspective with transformers.
\newblock In \emph{CVPR}, pages 6881--6890, 2021.

\bibitem[Zhu et~al.(2024)Zhu, Liao, Zhang, Wang, Liu, and Wang]{zhu2024vision}
Lianghui Zhu, Bencheng Liao, Qian Zhang, Xinlong Wang, Wenyu Liu, and Xinggang Wang.
\newblock Vision mamba: Efficient visual representation learning with bidirectional state space model.
\newblock \emph{arXiv preprint arXiv:2401.09417}, 2024.

\bibitem[Zhu et~al.(2017)Zhu, Xiong, Dai, Yuan, and Wei]{zhu2017deep}
Xizhou Zhu, Yuwen Xiong, Jifeng Dai, Lu Yuan, and Yichen Wei.
\newblock Deep feature flow for video recognition.
\newblock In \emph{CVPR}, pages 2349--2358, 2017.

\end{thebibliography}
}

\end{document}


\maketitlesupplementary
\appendix
\setcounter{figure}{0} 
\renewcommand{\thefigure}{S\arabic{figure}} 
\setcounter{table}{0} 
\renewcommand{\thetable}{S\arabic{table}}

This supplementary material provides more results that enhance and extend the findings presented in the main manuscript. Due to space constraints, certain details and experiments were omitted from the primary manuscript. Specifically, \cref{sec:additional_ablation}  presents more ablation studies that offer deeper insights into the proposed TV3S model. \cref{sec:updated_metric} details the latest performance results substantiating the efficacy of our proposed method through a more fair and refined training procedure. \cref{sec:additional_qual} showcases an expanded set of visual results demonstrating the segmentation capabilities of TV3S, alongside comparative analyses with additional models including MRCFA. 

\section{Additional Ablation Studies}
\label{sec:additional_ablation}

Following the main text, all ablation studies were conducted on the VSPW dataset using the MiT-B1 and Swin-T backbones, adhering to the same training and inference strategies outlined in the main text.

\para{Effect of spatial information extraction.}
To assess the effectiveness of our proposed TV3S architecture in extracting spatial information, we conducted experiments using single-frame inputs and compared the performance against baseline segmentation models and other video semantic segmentation (VSS) methods, as presented in \cref{supp:tab:1F_comparision}. While VSS methods are inherently designed for multi-frame processing, this evaluation isolates their ability to handle spatial features independently. For a fair comparison, we evaluated our model with and without the TV3S blocks, noting that our architecture can utilize the temporal blocks even when only one frame is provided. The results demonstrate that our model not only performs on par with the baseline when the TV3S blocks are excluded but also significantly outperforms it when the blocks are included.  In contrast, other VSS methods exhibit reduced performance in single-frame evaluations, reflecting their ability to partially adapt to single-frame inputs despite their multi-frame design. These findings indicate that our TV3S model effectively captures spatial information and maintains robust performance even without temporal context, showcasing its superiority in both spatial and spatiotemporal segmentation tasks.

\para{Effect of the number of TV3S blocks.} 
As detailed in the main text, the MiT-B1 backbone exhibited enhanced performance with an increasing number of TV3S blocks, achieving a mIoU of 40.0 and improved temporal consistency metrics (mVC\textsubscript{8} = 90.7, mVC\textsubscript{16} = 87.0) when utilizing four blocks, as shown in \cref{supp:tab:Eval_train_Block_full}. Extending this evaluation to the Swin-T backbone and maintaining a consistent framework, the Swin-T backbone attained a mIoU of 44.90 with four blocks, closely aligning with its peak performance of 45.11 achieved using two blocks. Additionally, temporal consistency metrics (mVC\textsubscript{8} = 88.0, mVC\textsubscript{16} = 83.5) remained stable across different block configurations. These findings indicate that, while the MiT-B1 backbone benefits significantly from an increased number of TV3S blocks, the Swin-T backbone maintains robust performance with a standardized four-block setup, underscoring the effectiveness of a unified framework for diverse backbones.

\begin{table}[!t]
  \centering
  \setlength{\tabcolsep}{4.8mm}
  \resizebox{\linewidth}{!}{
  \begin{tabular}{l|c|c|c}\toprule
    Methods & Backbones & mIoU$\uparrow$ & WIoU \\
    \midrule
    Segformer                            & MiT-B1        & 36.5          & 58.8\\
    CFFM                                & MiT-B1        & 37.1          &   59.0\\
    MRCFA                               & MiT-B1        & 37.0          & 58.8\\
    TV3S  (Ours)                        & MiT-B1        & 37.7          & 59.2\\ 
    TV3S  (+Blocks)                     & MiT-B1        & \textbf{38.6} & \textbf{60.3}\\ 
    Segformer    & MiT-B2        & 43.9          & 63.7\\
    CFFM                     & MiT-B2        & 43.6          & 63.3\\
    MRCFA                   & MiT-B2        & 43.4          & 63.5\\
    TV3S  (Ours)                        & MiT-B2        & 43.8          & 62.8\\ 
    TV3S  (+Blocks)                     & MiT-B2        & \textbf{44.9}          & \textbf{63.7}\\ 
    \midrule
    Segformer    & MiT-B5        & 48.9          & 65.1\\
    CFFM                     & MiT-B5        & 48.3          &  65.8\\
    MRCFA                   & MiT-B5        & 48.0          & 65.3\\
    TV3S  (Ours)                        & MiT-B5        & 48.9          & 66.0\\ 
    TV3S  (+Blocks)                     & MiT-B5        & \textbf{49.5}          & \textbf{66.4}\\ 
    \midrule
    Mask2Former       & Swin-T        & 41.2          &  62.6\\
    TV3S  (Ours)                        & Swin-T        & 42.8          & 62.4\\ 
    TV3S  (+Blocks)                     & Swin-T        & \textbf{43.8}          & \textbf{62.6}\\ 
    Mask2Former       & Swin-S        & 42.1          &  63.1\\
    TV3S  (Ours)                        & Swin-S        & 49.5          & 65.8\\ 
    TV3S  (+Blocks)                     & Swin-S        & \textbf{50.5}          & \textbf{66.2}\\ 
    \bottomrule
  \end{tabular}}
  \vspace{-6pt}
  \caption{Comparative effectiveness of models in extracting spatial information from single-frame inputs on the VSPW dataset, with our proposed method outperforming existing models.}
  \label{supp:tab:1F_comparision}
  \vspace{-.1in}
\end{table}

\begin{table*}[!t]
  \centering
  \setlength{\tabcolsep}{4.8mm}
  \resizebox{\linewidth}{!}{
  \begin{tabular}{l|c|c|c|c|c|c|c}\toprule
    Methods & Backbones & mIoU$\uparrow$ & mVC\textsubscript{8}$\uparrow$ & mVC\textsubscript{16}$\uparrow$ & GFLOPs$\downarrow$ & Params(M)$\downarrow$ & FPS$\uparrow$ \\
    \midrule
    Mask2Former  & R50 & 38.5 & 81.3 & 76.4 & 110.6 & 44.0 & 19.4 \\
    MPVSS        & R50 & 37.5 & 84.1 & 77.2 & 38.9 & 84.1 & 33.9\\
    Mask2Former  & R101 & 39.3 & 82.5 & 77.6 & 141.3 & 63.0 & 16.9\\
    MPVSS        & R101 & 38.8 & 84.8 & 79.6 & 45.1 & 103.1 & 32.3\\
    DeepLabv3+   & R101 & 34.7 & 83.2 & 78.2 & 379.0 & 62.7 & 9.2\\
    UperNet      & R101 & 36.5 & 82.6 & 76.1 & 403.6 & 83.2 & 16.0\\
    PSPNet       & R101 & 36.5 & 84.2 & 79.6 & 401.8 & 70.5 & 13.8\\
    OCRNet       & R101 & 36.7 & 84.0 & 79.0 & 361.7 & 58.1 & 14.3\\
    TCB          & R101 & 37.8 & 87.9 & 84.0 & 1692 & - & -\\
    ETC          & OCRNet & 37.5 & 84.1 & 79.1 & 361.7 & - & - \\
    Segformer    & MiT-B5 & 48.9 & 87.8 & 83.7 & 185.0 & 82.1 & 9.4\\
    CFFM         & MiT-B5 & 49.3 & 90.8 & 87.1 & 413.5 & 85.5 & 4.5\\
    MRCFA        & MiT-B5 & 49.9 & 90.9 & 87.4 & 373.0 & 84.5 & 5.0\\
    TV3S  (Ours) & MiT-B5 & \textbf{50.4} & \textbf{91.9} & \textbf{89.1} & 137.0& 85.6 & 14.0\\ 
    \bottomrule
  \end{tabular}}
  \vspace{-6pt}
  \caption{\textbf{Updated} quantitative comparison of our MiT-B5 model with existing methods on the VSPW dataset. Our model achieves the best balance among \textit{accuracy}, \textit{model complexity}, and \textit{operational speed}. FPS and FLOPs are calculated with an input resolution of 480 × 853.}
  \label{supp:tab:UpdatedVSPW_comparision}
  \vspace{-.1in}
\end{table*}

\begin{table}[!t]
    \centering
    \setlength{\tabcolsep}{3.3mm}
    \resizebox{\linewidth}{!}{
    \begin{tabular}{c|c|c|c|c} \toprule
        Backbones & TV3S Blocks & mIoU & mVC\textsubscript{8} & mVC\textsubscript{16} \\ 
        \midrule
        \multirow{4}{*}{MiT-B1} & 
                  1      & 38.4& 88.3& 83.7\\
                & 2      & 39.2& 89.5& 85.3\\
                & 3      & 39.6& 88.7& 84.2\\
                & \textbf{4}      & \textbf{40.0}& \textbf{90.7} & \textbf{87.0}  \\
        \midrule
        \multirow{4}{*}{Swin-T} & 
                  1      & 44.66  & 87.9  & 83.3  \\
                & \textbf{2}      & \textbf{45.11} & 88.4 & 83.9  \\
                & 3      & 44.41 & 88.3 & 83.8  \\
                & \textbf{4}      & \textbf{44.90} & 88.0 & 83.5  \\
        \bottomrule
    \end{tabular}}
    \vspace{-.1in}
    \caption{Performance metrics based on the number of TV3S blocks in the model.}
    \label{supp:tab:Eval_train_Block_full}
    \vspace{-.05in}
\end{table}

\para{Training with different temporal context.} 
We assessed the impact of varying the number of template frames during training on the MiT-B1 backbone variant of TV3S, as detailed in \cref{supp:tab:Eval_train_temporal}. Specifically, the model was trained with one ($\{I_{t-3}, I_{t}\}$), two ($\{I_{t-6}, I_{t-3}, I_{t}\}$), three ($\{I_{t-9}, I_{t-6}, I_{t-3}, I_{t}\}$) and five ($\{I_{t-15}, I_{t-12}, I_{t-9}, I_{t-6}, I_{t-3}, I_{t}\}$) template frames. The results indicate a clear improvement in visual consistency as the number of templates increases, showcasing the model’s enhanced ability to maintain temporal coherence, attributed to the specialized training methodology. However, while using five templates yielded the highest mVC values, the mIoU performance peaked with three templates, offering a balanced trade-off between segmentation accuracy and temporal consistency. Although further fine-tuning could refine the model for specific scenarios, the configuration with three templates is recommended for its optimal balance, aligning with findings from. This configuration ensures the model operates effectively within practical constraints while leveraging its temporal modeling strengths.

\begin{table}[!t]
    \centering
    \scriptsize
    \setlength{\tabcolsep}{3.3mm}
    \resizebox{\linewidth}{!}{
    \begin{tabular}{c|c|c|c} \toprule
        Templates No. & mIoU & mVC\textsubscript{8} & mVC\textsubscript{16} \\ 
        \midrule
        1      & 38.1& 90.3& 83.6\\
        2      & 37.6& 90.5& 84.3\\
        3      & \textbf{40.0}& 90.7 & 87.0  \\
        5      & 38.1& \textbf{91.2} & \textbf{88.0}  \\
        \bottomrule
    \end{tabular}}
    \vspace{-.1in}
    \caption{Evaluation based on the number of templates exposed during training.}
    \label{supp:tab:Eval_train_temporal}
    \vspace{-.05in}
\end{table}

\begin{table}[!t]
    \centering
    \renewcommand{\arraystretch}{1.} 
    \setlength{\tabcolsep}{5mm}
    \resizebox{\linewidth}{!}{
    \begin{tabular}{c|c|c|c}
        \toprule
        \multirow{2}{*}{Models} & \multicolumn{3}{c}{Evaluation (mIoU)} \\ \cmidrule{2-4} 
        & Bi & Bi+Embed & Direct \\ 
        \midrule
        1 TSS (No Shift) & 37.33 & 38.0 & 38.0\\ 
        TV3S (No Shift)  & 38.0  & 38.4 & 38.9 \\ 
        TV3S (Shift)     & \textbf{39.6}  & 37.6 & 39.5\\ 
        \bottomrule
    \end{tabular}}
    \vspace{-5pt}
    \caption{Implications of using bi-directional representation with embedding on the proposed architecture.}
    \label{supp:tab:Eval_internal_Bi}
    \vspace{-.1in}
\end{table}

\begin{table}[!t]
    \centering
    \scriptsize
    \setlength{\tabcolsep}{3.3mm}
    \resizebox{\linewidth}{!}{
    \begin{tabular}{c|c|c|c|c} \toprule
        Methods  &Backbones& mIoU & mVC\textsubscript{8} & mVC\textsubscript{16} \\ 
        \midrule
        VideoMamba&MiT-B1& 36.2& 83.9& 78.7\\
        TV3S&MiT-B1& 40.0& 90.7& 87.0\\
        \midrule
        MPVSS&Swin-B& 52.6& 89.5& 85.9\\
        MPVSS&Swin-L& 53.9& 89.6& 85.8\\
        TV3S&Swin-B& 53.0& 90.3 & 86.8  \\
        TV3S&Swin-L& 55.6& 90.7& 87.5\\
        \bottomrule
    \end{tabular}}
    \vspace{-.1in}
    \caption{Additional Experiments with VideoMamba as decoder and with bigger Swin Transformer backbones}
    \label{supp:tab:additional_exp}
    \vspace{-.05in}
\end{table}

\para{Applicability of Bi-directional Scanning.}
We investigated the use of bi-directional scanning, a technique prevalent in recent vision-based approaches utilizing mamba, in the MiT-B1 variant of TV3S (see \cref{supp:tab:Eval_internal_Bi}). This method involved scanning the encoded feature space in both directions, with or without adding embeddings during the scanning process, effectively doubling the computational load for the decoder. The experimental results indicated that incorporating bi-directional scanning did not enhance performance and, in some cases, led to degradation. We believe that this decline may be due to two factors: first, the implementation was conducted in a pixel-wise manner within the encoded feature space, differing from the patch-wise approach in the original mamba implementations; second, scanning the same feature space twice might disrupt the continuity of information, potentially hindering the model's ability to maintain performance. Consequently, these findings suggest that while bi-directional scanning is effective in certain contexts, its application as a decoder in the present architecture did not yield benefits and may require further methodological refinements.

\setlength\belowcaptionskip{+2ex}
\begin{figure*}[!t]
    \centering
    \includegraphics[width=\linewidth]{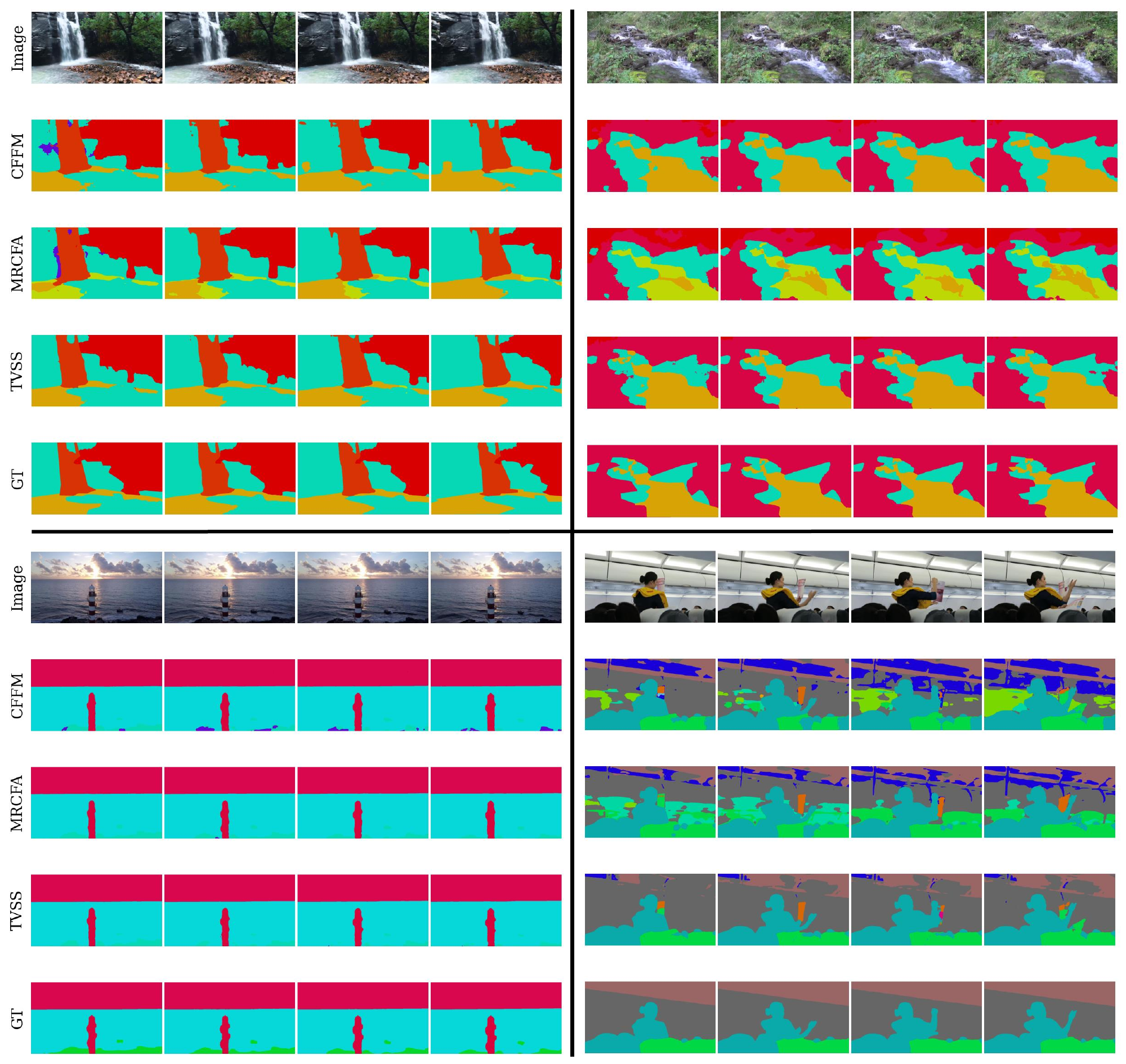}
    \vspace{-.20in}
    \caption{Additional examples showcasing the performance of the proposed TVSS architecture compared with other VSS methods, demonstrating visual consistency and accuracy.}
    \label{supp:fig:add_res_comp}
    \vspace{-.25in}
\end{figure*}

\para{Additional Experiments.}
Extended experiments were conducted during the rebuttal phase, which included testing VideoMamba and larger backbones of Swin, specifically its Swin-B and Swin-L variants, as tabulated in \cref{supp:tab:additional_exp}.  For the experiments with VideoMamba, we used the MiT-B1 backbone in conjunction with VideoMamba as the decoder. It was observed that VideoMamba only achieved a mean Intersection over Union (mIoU) of 36.24, while our TV3S framework achieved an mIoU of 40.0, thanks to its effective state propagation and shifted-window mechanism, making it ideal for dense prediction tasks.

As for the experiments involving larger backbones, it was noted that by directly extending the current framework without hyper-parameter tuning, we achieved mIoU scores that are better than the performance of MPVSS. This finding highlights the robustness of our approach and ensures fair comparisons with other methods.

\setlength\belowcaptionskip{+2ex}
\begin{figure*}[!t]
    \centering
    \includegraphics[width=\linewidth]{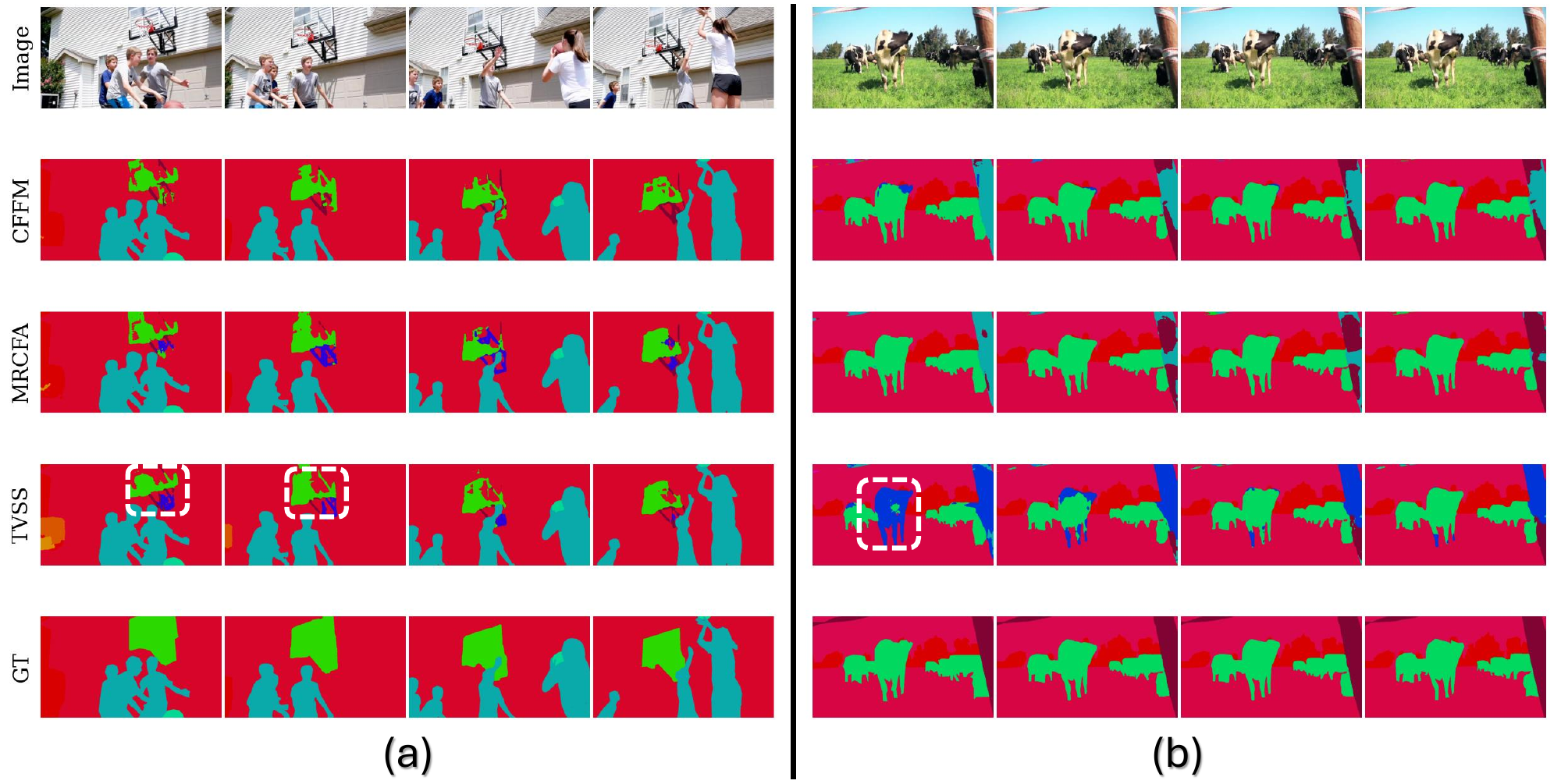}
    \caption{Failure cases of the proposed method: (a) errors in the presence of transparent objects and (b) initial segmentation errors propagating temporarily before being corrected.}
    \label{supp:fig:fail_res_comp}
    \vspace{-.1in}
\end{figure*}

\section{Updated Performance}
\label{sec:updated_metric}
Our initial training setup for the TV3S architecture, based on the \texttt{MMSegmentation} codebase, utilized two A100 NVIDIA GPUs with a batch size of 2 and trained the model for 160k iterations using three reference frames. This configuration resulted in strong temporal consistency metrics (mVC\textsubscript{8} and mVC\textsubscript{16}), achieving a good trade-off between computational efficiency and frames per second (FPS). However, compared to other video semantic segmentation (VSS) methods that were trained using four GPUs, our model was exposed to fewer data variants, potentially impacting its generalization capabilities.

To ensure a fairer comparison, we extended the training duration by an additional 80k iterations, totalling 240k iterations—a 50\% increase in training time. This adjustment compensates for the advantages other methods gain from using more GPUs, such as exposure to a wider variety of data and improved generalization. Concurrently, we halved the learning rate to 3e-5 from 6e-5  to maintain effective learning without overshooting, keeping the optimizer and learning rate scheduler configurations consistent.

Under this training setup, as shown in \cref{supp:tab:UpdatedVSPW_comparision} , our proposed TV3S architecture achieved state-of-the-art performance across all evaluated metrics, including mean Intersection over Union (mIoU) and temporal consistency metrics (mVC\textsubscript{8} and mVC\textsubscript{16}).

\begin{figure*}[!t]
    \centering
    \includegraphics[width=\linewidth]{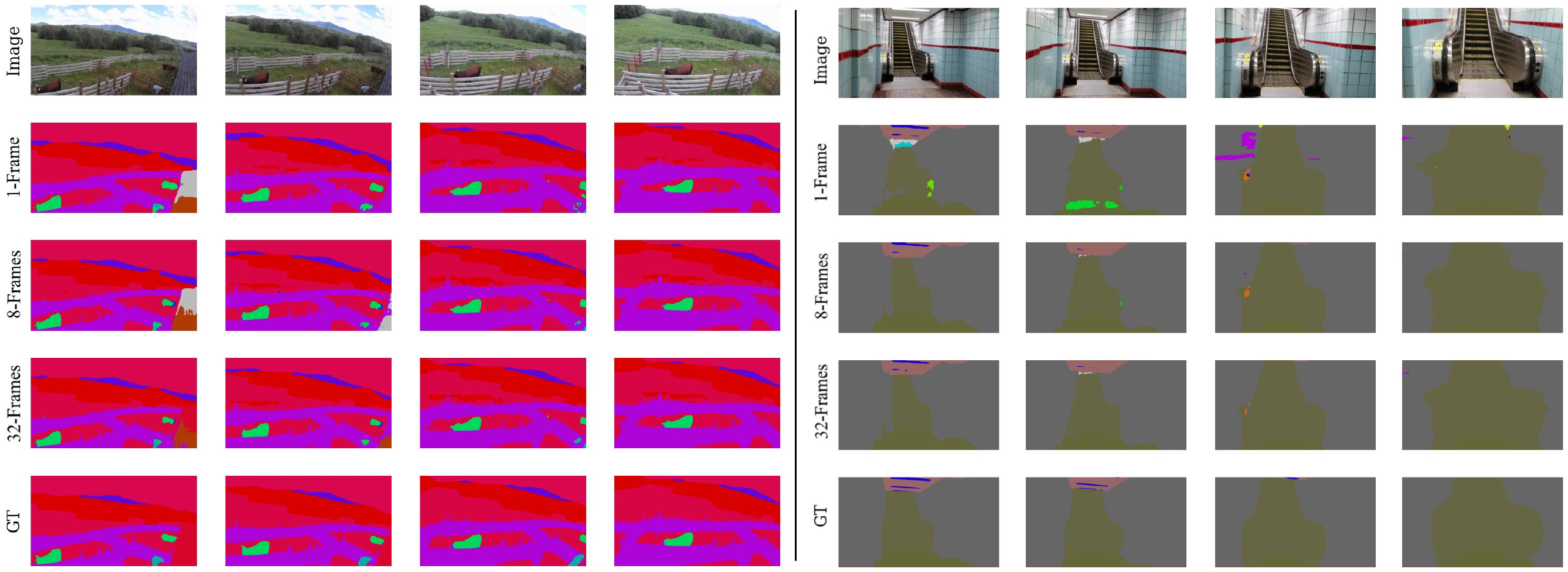}
    \vspace{-.20in}
    \caption{Visual comparison of segmentation results with 1, 8, and 32 exposed frames during inference.}
    \label{supp:fig:addinal_viz}
    \vspace{-.25in}
\end{figure*}

\section{Additional Qualitative Examples}
\label{sec:additional_qual}
In this section, we present qualitative examples to further demonstrate the effectiveness of the proposed TVSS architecture. As shown in \cref{supp:fig:add_res_comp}, the segmentation outputs from TVSS are compared with those from other state-of-the-art video semantic segmentation (VSS) methods. The examples illustrate how TVSS maintains good visual consistency across consecutive frames while achieving accurate segmentation. These results underline the advantages of the temporal state-sharing mechanism, which effectively propagates temporal information and reduces inconsistencies commonly observed in other methods. The visualizations in \cref{supp:fig:add_res_comp} provide a clear, comparative insight into how TVSS handles challenging scenarios, reinforcing the quantitative results discussed earlier.

\subsection{Success Cases} \label{ssec:success_qual} 

The proposed TVSS architecture excels in ensuring both stability and continuity in the segmentation process across frames, maintaining a high level of consistency even in dynamic and complex environments. The following examples demonstrate the architecture's ability to preserve these qualities in challenging visual sequences.

\para{(a) Temporal continuity and object consistency:} One of the standout features of TVSS is its ability to maintain temporal continuity. In the provided sequences, the model shows a consistent and stable segmentation of dynamic objects, such as waterfalls, people, or animals, across multiple frames. This is particularly evident in cases where objects remain in motion or where the background changes slightly, but the segmentation boundaries remain stable, offering a smooth transition between frames.

\para{(b) Robust segmentation in variable environments:} In more challenging scenes, including those with changing lighting or background complexity, TVSS continues to show visual stability. The segmentation boundaries are not only preserved, but also remain consistent across frames, regardless of the varying environmental conditions. The architecture's robustness to these changes ensures that even as new elements or disturbances appear, the model still provides coherent and unified segmentation results, reflecting its strong capacity to maintain accuracy over time.

These success cases underline the TVSS architecture's ability to offer consistent and continuous segmentation of objects, crucial for maintaining visual coherence across video sequences. The model's strength lies in its ability to handle the temporal aspect of visual data, ensuring that segmentation evolves seamlessly across frames without disruptions.

\subsection{Failure Cases}
\label{ssec:challenging_qual}
While the proposed TVSS architecture demonstrates robust performance across various scenarios, it is not without limitations. \cref{supp:fig:fail_res_comp} illustrates two primary challenging scenarios where the model encounters difficulties.

\para{(a) Transparent objects:}
The first set of failure cases involves the presence of transparent objects. Transparent materials often present ambiguous visual cues, making it challenging for segmentation models to accurately delineate boundaries and classify regions. In these instances, TVSS may misinterpret the transparency, leading to incorrect segmentation of the object or its background.

\para{(b) Error propagation from initial mis-classification:}
The second set of challenges pertains to the propagation of initial segmentation errors. When the model makes an initial misclassification in a frame, this error can propagate to subsequent frames due to the temporal state-sharing mechanism. Although TVSS is designed to leverage temporal information to enhance consistency, early mistakes can temporarily degrade segmentation accuracy until corrective learning occurs in subsequent frames.

These failure cases highlight areas for potential improvement, such as incorporating specialized modules for handling transparent materials and enhancing error correction mechanisms to mitigate the impact of initial misclassifications. Addressing these challenges will further strengthen the reliability and applicability of the TVSS architecture in diverse and complex environments.

\subsection{Additional Visualizations}

To qualitatively analyze segmentation consistency in videos and its effect based on the number of frames used during inference, we present \cref{supp:fig:addinal_viz}. Visual comparisons demonstrate that the results from using only one frame exhibit rough and fragmented segmentations. In contrast, predictions made using eight or thirty-two frames show smoother and more refined boundaries, closely resembling the ground truth (GT). This observation underscores the model's ability to effectively integrate temporal information, leading to better object delineation and improved segmentation boundaries. The enhanced consistency and quality of segmentation suggest that incorporating more frames enables the model to capture dynamic features and contextual information more effectively, particularly in challenging or ambiguous areas. This improvement can be attributed to the model's capacity to learn from the additional frames, resulting in a more accurate representation of the scene. This is especially apparent in complex or cluttered environments, where utilizing multiple frames significantly enhances the robustness and overall accuracy of the segmentation.

{
    \small
    \bibliographystyle{ieeenat_fullname}
    \bibliography{main}
}